\documentclass[slac_one,amsfonts,amssymb]{revtex4}
\usepackage{graphicx}
\usepackage{fancyhdr}
\renewcommand{\baselinestretch}{1.2}
\def\METs{{\mbox{$E\kern-0.57em\raise0.19ex\hbox{/}_{T}$}}}
\def\MET{{\mbox{$E\kern-0.57em\raise0.19ex\hbox{/}_{T}$}}\ }
\def\METc{{\mbox{$E\kern-0.57em\raise0.19ex\hbox{/}_{T}^{\it cal}$}}\ }

\def\ct{\mbox{$\cos\theta^*$}}

\def\ipb{{pb$^{-1}$\ }}
\def\itev{{TeV$^{-1}$}}

\begin{document}

\title{Collider Searches for Extra Dimensions}


\author{G. Landsberg\footnote{On behalf of the CDF and D\O\ Collaborations.}\footnote{E-mail: landsberg@hep.brown.edu.}}
\affiliation{Brown University, Department of Physics, 182 Hope St., Providence, RI 02912, USA
}

\begin{abstract}
Searches for extra spatial dimensions remain among the most popular 
new directions in our quest for physics beyond the Standard Model.
High-energy collider experiments of the current decade should be able
to find an ultimate answer to the question of their existence in a variety 
of models. Until the start of the LHC in a few years, the Tevatron will
remain the key player in this quest. In this paper, we review the most recent
results from the Tevatron on searches for large, TeV$^{-1}$-size, and 
Randall-Sundrum extra spatial dimensions, which have reached a new level of 
sensitivity and currently probe the parameter space beyond the existing constraints.
While no evidence for the existence of extra dimensions has been found so far,
an exciting discovery might be just steps away.
\end{abstract}

\maketitle

\thispagestyle{fancy}


\section{INTRODUCTION}

That our space might have a more complicated texture than convential three dimensions of infinite (or nearly infinite) size has been suggested a long time ago. The original idea perhaps goes back as far as Bernhard Riemann's famous Habilitation lecture~\cite{Riemann} in which he introduced the concepts of what are now known as the Riemannian space and curvature tensor. It was not until the the XXth century that these novel ideas have been first employed by physicists: with Albert Einstein's general relativity (GR) theory~\cite{Einstein} based on the Riemann's concept of space, and Theodore Kaluza and Oscar Klein's interesting, albeit unsuccessful attempt to unify GR with electromagnetism (EM) by introducing an additional spatial dimension of a finite size~\cite{KK} (in what follows referred to as a {\it compact dimension\/}). In 1970-ies the concept of compact dimensions has been utilized by string theory, that requires additional six or seven very small spatial dimensions (usually with the size $\sim 10^{-35}$ m) for a self-consistent description of quantum gravity. Unfortunately, such tiny extra dimensions (ED) cannot be probed in any existing or foreseen experiments as the ``resolution'' necessary to penetrate such small distances would require a center-of-mass energy $\sim 10^{13}$~TeV, i.e. far beyond the range spanned by either natural or man-made beams. Consequently, these extra dimensions remain more of a mathematical concept than a falsifiable physics prediction, at least for now.

For a particle propagating in the extra dimensions the boundary conditions allow only for certain eigenvalues of energy (cf. classical ``particle in a box'' problem). These quantized energy states look like a tower of massive particles from the point of view of a 3-dimensional (3D) observer and are referred to as Kaluza-Klein (KK) modes or excitations of a particle. Typical energy spacing\footnote{Here and in what follows we use ``natural'' units $\hbar = c = 1$.} between the KK modes is $\sim 1/R$. As the following discussion will reveal, these KK-modes can modify the conventional picture of physical world in a most profound way.

It was not until 1998 that the idea of extra dimensions in space has been used as a possible solution of the one of the most embarassing problems of the Standard Model (SM) of particle physics -- the hierarchy problem. The essence of this problem can be captured in a single question: why is gravity (at least as we observe it) some $10^{38}$ orders of magnitude weaker than other forces of Nature? This mysterious fact is responsible for a tremendous hierarchy of energy scales: from the scale of electroweak (EW) symmetry breaking ($M_{\rm EW} \sim 1$ TeV) to the energies at which gravity is expected to become as strong as other known forces (strong, EM, and weak), known as Planck scale or $M_{\rm Pl} = 1/\sqrt{G_N} = \sim 10^{16}$ TeV, where $G_N$ is the Newton's coupling constant.

Why is the hierarchy problem so annoying? The reason is that in our everyday experience, large hierarchies tend to collapse, unless they are supported by some intermediate layers or by a certain amount of fine tuning of the initial condtions. E.g., it is mechanically possible to balance a billiard cue on its point, but the amount of fine tuning required for such a delicate balance is too demanding to be of a practical use. Indeed, walking in a room and seeing a cue standing steadily on its end would be too odd not to suspect a some kind of hidden support. Similarly, making the SM work for all energies up to the Planck scale would require tremendous amount of fine tuning of its parameters~-- to a precision of $\sim (M_{\rm Pl}/M_{\rm EW})^2 \sim 10^{-32}$. A natural reaction to this observation is to conclude that the SM is not complete and certain new physics has to exhibit itself at the energies of the order of the EW scale, thus ``supporting'' the hierarchy with an extra layer.

Nevertheless, it's worth noting that large amounts of fine tuning, while unnatural, are not prohibited by any fundamental principle. The fact that large hierarchies tend to collapse is merely empirical. While many large hierarchies we know have indeed collapsed (e.g., human towers, once they have too many layers, countless monarchies, the Soviet empire), some others are surprisingly stable (e.g., Egyptian pyramids, Eiffel tower, George W. Bush's government). Note also that certain examples of fine tuning have been observed in Nature:
\begin{itemize}
\item One of the best known examples of fine tuning is the fact that the apparent angular size of the moon is the same as the angular size of the Sun within 2.5\%~-- a mere coincidence to which we owe such a spectacular sight as solar eclipse.
\item Somewhat less abused example is the infamous Florida 2000 presidential election recount with the official result of 2,913,321 Republican vs. 2,913,144 Democratic votes, with the ratio equal to 1.000061, i.e. fine-tuned to one with the precision of 0.006\%.
\item Yet less known example\footnote{This happens to be not just a mere coincidence, but actually an example of a certain ``hidden'' principle that guarantees large amount of fine tuning. In this example, it turns out that in the base-$N$ numerical system the ratio of the number composed of digits $N-1$ through 1 in the decreasing order to the number obtained from the same digits, placed in the increasing order, is equal to $N - 2$ with the precision assymptotically approaching $10^{-N}$. Playing with a hexadecimal calculator could easily reveal this via the observation that the ratio of FEDCBA987654321 to 123456789ABCDEF is equal to 14.000000000000000183, i.e. 14 with the precision of $1.3\times 10^{-17}$. Whether the $10^{-38}$ precision needed to fine-tune the SM could be a result of a similarly hidden principle is yet to be found out.} is that the ratio of 987654321 to 123456789 is equal to 8.000000073, i.e. eight with the precision of $9.1\times 10^{-9}$.
\end{itemize}

\section{EXTRA DIMENSIONS AND THE HIERARCHY PROBLEM}

The first attempt to solve the hierarchy problem using extra spatial dimensions was made by Nima Arkani-Hamed, Savas Dimopoulos, and Georgi Dvali (ADD)~\cite{ADD}. The idea was drastically different from the common utilization of extra dimensions in string theory. In the ADD paradigm extra spatial dimensions are used to create an extra volume, which is sufficiently large to dilute gravity, so that it appears weak to a 3D-observer. To realize this the extra dimensions have to be many orders of magnitude larger than those in string theory. In order for such ``large'' extra dimensions not to violate any constraints from atomic physics and other experimental data, all other particles except for the graviton must not be allowed to ``feel'' the extra space. Fortunately, modern quantum field theory allows to confine spin 0, 1/2, and 1 particles to a subset of a multidimensional space, or a ``brane.'' At the same time, the graviton, being a spin-2 particle, is not confined to the brane and thus penetrates the entire (``bulk'') space. The solution to the hierarchy problem (or rather its ``removal'') then becomes straightforward: if the Planck scale in the multidimensional space is of the order of the EW scale, then there is no large hierarchy and no fine-tuning problem to deal with, as non-trivial physics really ``ends'' at the energies $\sim 1$~TeV. Gravity appears weak to a 3D-observer merely because it spans the entire bulk space; therefore the gravitons spend very little time in the vicinity of our brane. Gravity in 3D is literally diluted by the enormous volume of extra space.

The size of large extra dimensions ($R$) is fixed by their number ($n$) and the {\it fundamental\/} Planck scale in the (4$+n$)-dimensional space-time ($M_D$). By applying Gauss's law, one finds~\cite{ADD,GRW}:
\begin{equation}
        M^2_{\rm Pl} = 8\pi M_D^{n+2}\, R^n.
\end{equation}
(Here for simplicity we assumed that all $n$ dimensions have the same size $R$ and are compactified on a torus.)

If one requires $M_D \sim 1$~TeV and a single extra dimension, its size has to be of the order of the radius of the solar system; however, already for two ED their size is only $\sim 1$~mm; for three ED it is $\sim 1$~nm, i.e. similar to the size of an atom; and for larger number of ED it further decreases to subatomic sizes and reaches $\sim 1$~fm (the size of a proton) for seven ED.

A direct consequence of these compact extra dimensions in which gravity is allowed to propagate, is a modification of Newton's law at the distances comparable with the size of ED: the gravitational potential would fall of as $1/r^{n+1}$ for $r \lesssim R$. This clearly rules out the possibility of a single ED, as the very existence of our solar system requires the potential to fall of as $1/r$ at the distances comparable with the size of planetary orbits. However, it turns out that as of 1998 any higher number of ED have not been ruled out by any gravitational measurements, as Newton's law has not been tested to the distances smaller than about 1~mm. Amazing as it is, large ED with the size as macroscopic as 1~mm were perfectly consistent with the host of experimental measurements at the time when the ADD idea has appeared!

Note that the solution to the hierarchy problem in the ADD paradigm is incomplete. In a sense, the hierarchy of energy scales is traded for the hierarchy of distances from a ``natural'' electroweak symmetry breaking (EWSB) range of 1 TeV$^{-1} \sim 10^{-19}$~m to the much larger size of the compact extra dimensions. Nevertheless, this hierarchy could be significantly smaller than the hierarchy of energy scales or perhaps stabilized by some topological means. Strictly speaking, the idea of large extra dimensions is really a paradigm which can be used to build more or less realistic models, rather than a model by itself.

Almost simultaneously with the ADD paradigm a very different low-energy utilization of the idea of compact extra dimensions has been proposed by Keith Dienes, Emilian Dudas, and Tony Gherghetta~\cite{DDG}. In their model, additional dimension(s) of the ``natural'' EWSB size of $R \sim 1 $~TeV$^{-1}$~\cite{itev} are added to the SM to allow for a low-energy unification of gauge forces. In conventional SM and its popular extensions, such as supersymmetry, gauge couplings run logarithmically with energy, which is a direct consequence of the renormalization group evolution (RGE) equations. Given the values of the strong, EM, and weak couplings at low energies, all three couplings are expected to ``unify'' (i.e., reach the same strength) at the energy $\sim 10^{13}$~TeV, know as the Grand Unification Theory (GUT) scale. However, if one allows gauge bosons responsible for strong, EM, and weak interactions to propagate in extra dimensions, the RGE equations would change. Namely, once the energy is sufficient to excite KK modes of gauge bosons (i.e. $\sim 1/R \sim 1$~TeV), running of the couplings is proportional to a certain power of energy, rather than its logarithm. Thus, the unification of all three couplings can be achieved at much lower energies than the GUT scale, possibly as low as 10-100 TeV~\cite{DDG}.

An obvious deficiency of this model is that it does not incorporate gravity and thus does not explain its weakness relative to other forces. Nevertheless, it removes another hierarchy of a comparable size -- the hierarchy between the EWSB and GUT scales.\footnote{Since we don't understand gravity at the quantum level, it's not clear what would happen to the running of the gravitational coupling if a graviton is allowed to propagate in \itev\ ED.}

In 1999, Lisa Randall and Raman Sundrum offered a rigorous solution~\cite{RS} of the hierarchy problem by adding a single extra dimension (with the size that can range anywhere from $\sim 1/M_{\rm Pl}$ virtually to infinity) with a non-Euclidean, warped metric. They used the Anti-deSitter (AdS) metric (i.e. that of a space with a constant negative curvature) $ds^2 = \exp(-2kR|\varphi|)\eta_{\mu\nu}dx^\mu dx^\nu - R^2d\varphi^2$, where $0 \le \varphi < 2\pi$ is the coordinate along the extra spatial dimension of radius $R$, $k$ is the curvature of the AdS space (warp factor), $x^\mu$ are the convential (3+1)-space-time coordinates, and $\eta^{\mu\nu}$ is the metric of the Minkowski space-time. A (3+1)-dimensional brane with positive tension is put at $\varphi = 0$. If gravity originates on this brane (Planck brane), the wave function of the graviton has a peculiar feature that it is exponentially suppressed away from the brane in the direction of the extra dimension. If all the SM fields were confined to the Planck brane, one would not have seen any low-energy effects in this model. However, if a second brane with negative tension (SM brane) is put at $\varphi = \pi$, than the $M_{\rm Pl}$-size operators on the Planck brane would result in low-energy effects on the SM brane with the typical scale of $\Lambda_\pi = \overline{M}_{\rm Pl}\exp(-k\pi R)$, where $\overline{M}_{\rm Pl} \equiv M_{\rm Pl}/\sqrt{8\pi}$ is the reduced Planck mass. If the SM fields are confined to the SM brane, the hierarchy problem is solved for $\Lambda_\pi \sim 1$~TeV, which can be achieved with a little amount of fine tuning by requiring $kR \sim 10$. Since the only fundamental scale in this model is $M_{\rm Pl}$, the hierarchy problem is solved naturally for $R \sim 1/M_{\rm Pl}$. It is even possible to stabilize the Planck and SM branes at $\varphi = 0$ and $\pi$, respectively, by putting them in the fixed points of an orbifold in the AdS space.

In the simplest Randall-Sundrum (RS) model~\cite{RS}, gravitons are the only particles propagating in the AdS space. Several extensions of this model exist~\cite{RSext}, where various other particles are allowed to propagate in the bulk space, leading to rich low-energy phenomenology.

Another interesting feature of the RS model is the existence of a new, Higgs-like scalar particle, the radion which is a modulus field describing relative motion of the two branes~\cite{GW}. Radion phenomenology has been a subject of intense theoretical studies in the past few years~\cite{radion}. Particularly interesting is the Higgs-radion mixing, which would modify the branching fractions of the standard model Higgs, thus affecting the sensitivity of collider searches for the Higgs boson. Similar to the Higgs searches, high energy and integrated luminosity are the keys to the radion searches at colliders. However, in the allowed and theoretically preferred parameter space of the RS model it is very unlikely that the radion can be observed at the Tevatron, so a more detailed discussion of its phenomenology goes beyond the scope of this paper.

The most ``democratic'' of the ED models has been suggested in 2000 by Thomas Appelquist, Hsin-Chia Cheng, and Bogdan Dobrescu and is referred to as universal extra dimensions (UED)~\cite{UED}. In this model, all the SM particles are allowed to propagate in the bulk space of several extra dimensions of a TeV$^{-1}$ size. A special quantum number, the Kaluza-Klein parity, is conserved in this model: $n_1 \pm n_2 \pm n_3 \pm ... \pm n_N = const$, where $n_i$ is the Kaluza-Klein mode for the $i$-th particle of a certain multiparticle state. One of the consequences of this conservation law is that if one produces a final state of $N$ particles in collisions of the 0-th KK modes (e.g., at colliders), the Kaluza-Klein parity of the final state has to remain zero, which implies that one can not produce a single excited KK mode in such a collision. Consequently, as in SUSY with $R$-parity conservation, excited KK modes must be produced in pairs, which lifts many would-be constraints on the UED model. Another consequence is that the lightest KK excitation (LKKE)~-- that of a photon~-- must be stable and provides a suitable candidate for the cold dark matter~\cite{KKDM}. Radiative correctons to the KK spectrum of SM particles break mass degeneracy between KK excitations of various particles and result in SUSY-like cascade decays terminated with a pair of LKKE's in the final state, which could be even confused for a SUSY spectrum at the LHC and future colliders~\cite{KKSUSY}. While UED offer a number of interesting low-energy effects, there are no results on the UED searches at the Tevatron as of yet. Consequently, we won't discuss this model in any more detail here.

To summarize, various new models with extra dimensions can solve some of the deficiencies of the SM. While inspired by string theory, none of these models are based on it. Note that in principle several of the models can be realized in Nature simultaneously. For example, one could imagine a universe with six extra dimensions, three of which are large, one has a \itev\ size, and two others are of a very small radius expected naturally in string theory. Thus, string theory, low energy GUT, and TeV-scale gravity might in principle all coexist!

Finally, in 2002 yet another class of models based on ED has been suggested by Nima Arkani-Hamed, Andy Cohen, and Howard Georgi. In these models the concept of extra dimensions is used to build a multi-dimensional theory, which at the end of the day looks like a 4-dimensional one~\cite{ACG}. 
There exist interesting implementations of this class of models~\cite{LH}, in which the hierarchy problem is solved by introducing an additional scale $\sim 100$ TeV and a light Higgs boson (little Higgs) responsible for EWSB. Quadratic divergencies of the SM are cancelled up to this new scale by additional vector-like quarks. These models predict moderate low-energy phenomenology and typically require a few additional particles with masses $\sim 1$~TeV, particularly an extra neutral massive gauge boson $Z_H$ with the properties similar to those of the $Z$.

\section{PROBING EXTRA DIMENSIONS IN THE LAB}

The most experimentally-interesting feature of recent ED models is rich low-energy phenomenology that originates from the KK spectrum of various particles propagating in ED. In what follows we compare the KK spectrum observed in the ADD, \itev, and RS models and discuss various experimental constraints on model parameters.

\subsection{Kaluza-Klein Spectrum and Low-Energy Phenomenology}

In the ADD model, the only particle that propagates in ED and acquires KK modes is the graviton. Given the size of ED ($\sim 10^{-3}$--$10^{-15}$~m), energy spacing between the KK excitations of the graviton is $\sim 1$ meV -- 100~MeV. Consequently, the adjacent modes are hard to resolve and in most of the experiments the KK spectrum would appear continuous, from zero to a certain ultraviolet cutoff above which quantum gravity effects would modify this semi-classical picture. Since the fundamental Planck scale in the ADD model is $M_D \sim 1$~TeV, it's natural to expect this cutoff, $M_S$, to be of the same order: $M_S \sim M_D$. While each KK mode couples to the energy-momentum tensor with gravitational strength $G_N$, the sheer number of available KK modes is sufficient to enhance gravitational attraction tremendously. For example, in a particle collision with the energy $\sim 1$~TeV as many as $10^4$ KK modes along each extra dimension are accessible, even in the case of the smallest radius of these ED, $\sim 1$ fm. Given that this size corresponds to $n = 7$, a total of $10^{4n} = 10^{28}$ modes can be excited, thus magnifying gravitational interaction by this enormous factor.

In the \itev\ model with a single extra dimension of the size $R$, the 0-th KK mode of a gauge boson is the SM particle of mass $M_V$. This mass can be either $0$ (photon, gluon) or $>0$ ($W$, $Z$). The mass of the $i$-th KK mode is given by the following equation:
\begin{equation}
\label{eq:massKK}
	M_i = \sqrt{M_V^2 + i^2/R^2}.
\end{equation}
For the compactification scale $M_C \equiv 1/R \sim 1$~TeV, $1/R \gg M_V$ and hence KK modes of all gauge bosons are nearly degenerate. Consequently, the mass of the $i$-th mode is approximately equal to $i M_C$ for any $i > 0$. The couplings of all the KK modes of gauge bosons with $i > 0$ to the SM fields are $\sqrt{2}$ greater than the couplings of the corresponding SM bosons. This factor comes from the normalization of the KK modes (due to the fact that the negative values of $i$ are not considered and only the positive ones are kept in the effective Lagrangian). While the graviton, which is presumably free to propagate in the entire bulk space, would also acquire KK modes with the energy spacing $M_C$, these modes are sufficiently sparse not to affect graviational interaction by a visible amount. For example, if the energy of an interaction is as high as 100 TeV, the gravitational interaction is increased just by two orders of magnitude, i.e. imperceptibly small. The KK tower of excited gauge bosons is expected to be truncated at the masses of the order of the new GUT scale (i.e., $\sim 100$~TeV), where new physics (e.g., brane dynamics) should take over.

Finally, in the Randall-Sundrum model the 0-th KK mode of the graviton remains massless, while the higher modes have masses spaced as subsequent zeros of the Bessel's function $J_1$. Such an ``exotic'' solution comes from the fact that the graviton is confined to a ``box'' with a peculiar warped metric. On the other hand, an observation of just two excitations of the KK graviton mith the ratio of the masses being at a zero of $J_1$ would be a unique smoking-gun signature of warped extra dimensions. If the mass of the first excited mode of the graviton is $M_1$, the subsequent excitations would have masses of 1.83$M_1$, 2.66$M_1$, 3.48$M_1$, 4.30$M_1$, etc. The spacing between the subsequent KK excitations decreases at high masses. The 0-th excitation is coupled to the energy-momentum tensor with gravitational strength $G_N$, while each of the higher excitations couples with the strength $\sim 1/\Lambda_\pi^2$, i.e. comparable with the EW coupling strength. The excited modes are presumably truncated at masses $\sim M_{\rm Pl}$, where brane excitations would modify the discussed behavior.

\subsection{Direct Gravity Measurements}

The most straightforward observable effect of the large ED is the modification of Newton's gravitational attraction law at very short distances. Current and forseen techniques only allow these measurements to probe extra dimensions as small as $\sim 50~\mu$m, which limits their applicability to the case of $n = 2$, assuming that all ED have similar sizes. However, since the relationship between the fundamental and apparent Planck scales depends only on the total volume of extra space, and not on the shape of extra dimensions, it is possible that even for higher number of extra dimensions, one of them could be macroscopic, while the other ones are much smaller. Since gravity measurements are sensitive to the largest extra dimension, they can probe such asymmetric cases even for $n > 2$. Unfortunately, gravity measurements at short distances offer no sensitivity to the \itev\ and RS models.

Currently, the best limit on the size of extra dimensions in the ADD model from gravity measurements comes from the E\"ot-Washington group~\cite{Adelberger} and constrains the largest ED size to $R < 0.16$~mm at the 95\% confidence level (CL), which corresponds to $M_D > 1.7$~TeV in the case of two ED of equal size. For further discussion of the ``tabletop'' gravitational experiments, see Sylvia Smullin's talk at SSI `04~\cite{Smullin} or Ref.~\cite{Price}.

\subsection{Constraints from Astrophysics and Cosmology}

Certain constraints on the ADD model come from astrophysical observations and cosmology. For example, KK graviton ($G_{\rm KK}$) emission could serve as an alternative cooling mechanism for the supernovae, thus suppressing the dominant SM mechanism -- the neutrino emission. The observation of a handful of neutrinos from the SN1987A explosion by the IMB and Kamiokande detectors allowed to put a constraint of $M_D > 25$--30 TeV for $n = 2$ and 2--4 TeV for $n = 3$~\cite{astro}. For any higher number of ED limits on $M_D$ are below 1 TeV. Other constraints come from non-observation of a distortion of cosmic diffuse gamma-radiation spectrum expected from the decays of KK gravitons into pairs of photons~\cite{CDG} and from the lack of photon emission from the decays of gravitons trapped in the vicinity of neutron stars formed in the process of supernovae collapse~\cite{SN}. A significant excess of KK gravitons in early universe would result in its overclosure or early matter dominance, which also leads to constraints~\cite{cosmo} on $M_D$.

All these limits have similar features: they are typically rather strong (up to 1700 TeV!) for $n = 2$, moderate (a few TeV) for $n = 3$, and rather weak for $n > 3$. The other feature of these constraints is that all of the underlying calculations are based on a number of assumptions, so the results are reliable only as an order of magnitude estimate. (It is not unusual for different calculations based on the same data to differ by a factor of 2--3.) Nevertheless, $n = 2$ case seems to be largely disfavored by the host of the astrophysical and cosmological constraints.

None of the astrophysical or cosmological arguments result in any important limits on the \itev\ or RS models. Further review of constraints from astrophysics and cosmology can be found in Ref.~\cite{HS}.

\subsection{Collider Constraints on Extra Dimensions}

High-energy colliders are the finest microscopes invented by the human kind. The discovery of the top quark at the Fermilab Tevatron~\cite{top} demonstrated that distances as small as $\sim 10^{-18}$~m can be probed effectively. Hence, colliders remain the ultimate tool in probing extra dimensions in all the discussed models, for virtually any value of $n$.

\begin{figure}[hb]
\includegraphics[width=4.5in]{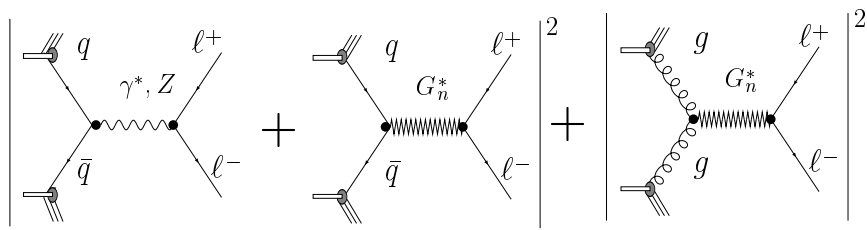}
\vspace{-0.2in}
\caption{Feynman diagrams for the modified DY production in the presence of large extra dimensions. From Ref.~\protect\cite{Gupta}.}
\label{fig:DY}
\end{figure}

Since KK gravitons couple to the energy-momentum tensor, they can be added to any vertex or line of a SM Feynman diagram. Consequently, to probe the ADD model at colliders one could look for direct emission of KK gravitons, e.g. in the $q \bar q \to g + G_{\rm KK}$ process, which result in a single jet (monojet) in the observable final state. The experimental signature for monojets is an apparent transverse momentum non-conservation and an overall enhancement of the tail of the jet transverse energy spectrum. Another type of effect observable in high-energy collisions is an enhancement of Drell-Yan (DY) or diboson spectrum at high invariant masses due to additional diagrams involving virtual KK graviton exchange. Typically, these diagrams interfere with their SM counterparts, as shown in Fig.~\ref{fig:DY} for the case of DY production.

Searches for virtual graviton effects are complementary to those for direct graviton emission, since the former depend on the ultraviolet cutoff of the KK spectrum, $M_S$, while the latter depends directly on the fundamental Planck scale $M_D$. While both scales are expected to be of the same order, it is quite possible that $M_S$ is somewhat lower than $M_D$; thus the effects of extra dimensions might be detected in virtual graviton exchange before they are observed in direct emission. 

While direct graviton emission cross section is well defined~\cite{Peskin,GRW}, the cross section for virtual graviton exchange depends on a particular representation of the interaction Lagrangian and the definition of the ultraviolet cutoff on the KK modes. Three such representations have appeared nearly simultaneously~\cite{Hewett,GRW,HLZ}. In all of them, the effects of ED are parameterized via a single variable $\eta_G = {\cal F}/M_S^4$, where ${\cal F}$ is a dimensionless parameter of order one reflecting the dependence of virtual $G_{\rm KK}$ exchange on the number of extra dimensions, and $M_S$ is the ultraviolet cutoff. Different formalisms use different definitions of ${\cal F}$, which result in different definitions of $M_S$:
\begin{eqnarray}
        {\cal F} & = & 1, \mbox{~(GRW~\cite{GRW});} \label{eq1}\\
        {\cal F} & = & \left\{ \begin{array}{ll} 
         \log\left( \frac{M_S^2}{M^2} \right), & n = 2 \\
           \frac{2}{n-2}, & n > 2
           \end{array} \right. , \mbox{~(HLZ~\cite{HLZ});}\\ \label{eq2}
        {\cal F} & = & \frac{2\lambda}{\pi} = \pm\frac{2}{\pi}, \mbox{~(Hewett~\cite{Hewett}).} 
        \label{eq3}
\end{eqnarray}

Note that ${\cal F}$ depends explicitly on $n$ only within the HLZ formalism. In both the GRW and HLZ formalisms gravity effects interfere construtively with the SM diagrams. In Hewett's convention the sign of intereference is not known, and the interference effects are parameterized via a parameter $\lambda$ of order one, which is usually taken to be either $+1$ (constructive interference) or $-1$ (destructive interference). The parameter $\eta_G$ has units of TeV$^{-4}$ if $M_S$ is expressed in TeV, and describes the strength of gravity in the presence of large ED. The differential or total cross section in the presence of virtual graviton exchange can be parameterized as:
\begin{equation}
	\sigma_{\rm tot} = \sigma_{\rm SM} + \eta_G \sigma_{\rm int} + \eta_G^2 \sigma_{G}, \label{eq:virtual}
\end{equation}
where $\sigma_{\rm SM}$ is the SM cross section for the process under study and $\sigma_{\rm int}$, $\sigma_G$ are the interference and direct graviton effects, respectively.

The most dramatic and exciting observable effect of large ED is copious production of black holes at future accelerators with the c.o.m. energy exceeding the fundamental Planck scale (e.g., at the LHC)~\cite{BH} and in ultra-high-energy cosmic-ray interactions~\cite{BHCR}. This topic goes beyond the Tevatron reach and is covered in separate reviews~\cite{BHreview}.

LEP experiments have pioneered searches for extra dimensions at colliders in both direct graviton emission in the $e^+ e^- \to \gamma/Z + G_{\rm KK}$ mode and via virtual graviton effects in DY production of various fermion pairs, as well as in the diboson production. The best sensitvity is achieved in the $e^+ e^- \to \gamma G_{\rm KK}$ and in the $e^+ e^- \to e^+ e^-, \gamma\gamma$ modes. The preliminary individual and combined LEP limits from direct graviton emission are listed in Table~\ref{table:LEP-direct}~\cite{LEP-direct}. Here and in what follows, all limits are quoted at the 95\% CL.

\begin{table}[hbt]%[H] add [H] placement to break table across pages
\caption{\label{table:LEP-direct}
Individual and combined 95\% CL limits on the fundamental Planck scale $M_D$ in the large ED model (in TeV) from the LEP experiments. OPAL limits are based only on the 1998 $\sqrt{s} = 189$~GeV data and have not been used in the combination.}
\begin{tabular}{lccccc}
% Lines of table here ending with \\
\hline
Experiment 	& $n=2$ & $n=3$ & $n=4$ & $n=5$ & $n=6$  \\
\hline
ALEPH		& 1.26	& 0.95	& 0.77  & 0.65  & 0.57   \\
DELPHI		& 1.31  & 1.02  & 0.82  & 0.67  & 0.58   \\
L3		& 1.50  & 1.14  & 0.91  & 0.76  & 0.65   \\
OPAL		& 1.09  & 0.86  & 0.71  & 0.61  & 0.53   \\
\hline
Combined (ADL)	& 1.60  & 1.20  & 0.94  & 0.77  & 0.66   \\
\hline
\end{tabular}
\end{table}

The combined LEP limit on $M_S$ from difermion and diphoton production in Hewett's convention is 0.95 TeV (destructive interference) or 1.14 TeV (constructive interference)~\cite{Bourilkov}.

Both the CDF and D\O\ Collaborations at the Fermilab Tevatron $p\bar p$ collider sought for large ED in Run I. D\O\ has pioneered searches for virtual graviton effects in the dielectron and diphoton channels at hadron colliders~\cite{D0-diem}, as well as searches in the challenging monojet channel~\cite{D0-monojet} plagued by copious instrumental backgrounds from jet mismeasurement and cosmics. CDF has pioneered searches in the monophoton channel at hadron colliders~\cite{CDF-monophoton} and also accomplished a monojet analysis~\cite{CDF-monojet}. D\O\ Run I limits on virtual graviton effects in Hewett's convention are $M_S > 1.0$ and 1.1 TeV for constructive and destructive interference, respectively. (These limits are the tightest limits on $M_S$ coming from a single experiment.)

The results of the monojet and monophoton searches in Run I are summarized in Table~\ref{table:RunI}. The unknown next-to-leading order (NLO) effects are parameterized via a constant $K$-factor and the limits are given at the leading order (LO, $K = 1$) and for $K = 1.3$, typical for many processes at the Tevatron.

\begin{table}[hbt]%[H] add [H] placement to break table across pages
\caption{\label{table:RunI}
Individual 95\% CL lower limits on the fundamental Planck scale $M_D$ (in TeV) in the ADD model from the CDF and D\O\ experiments. Ordering of the results is chronological.}
\begin{tabular}{lccccccc}
% Lines of table here ending with \\
\hline
Experiment and channel 	& $n=2$ & $n=3$ & $n=4$ & $n=5$ & $n=6$ & $n=7$ & $n=8$ \\
\hline
CDF monophotons, $K = 1$&       &       & 0.549 &       & 0.581 &       & 0.602 \\
\hline
D\O\ monojets, $K = 1$	& 0.89	& 0.73	& 0.68  & 0.64  & 0.63  & 0.62 \\
D\O\ monojets, $K = 1.3$& 1.99  & 0.80  & 0.73  & 0.66  & 0.65  & 0.63 \\
\hline
CDF monojets, $K=1$	& 1.00  &       & 0.77  &       & 0.71 \\
CDF monojets, $K=1.3$	& 1.06  &       & 0.80  &       & 0.73 \\
\hline
\end{tabular}
\end{table}

HERA experiments have also looked at the effects of the $t$-channel exchange of virtual gravitons in the $e^\pm p \to e^\pm p$ data and set limits on the ultraviolet cutoff $M_S$ of 0.82 and 0.79 TeV in Hewett's convention for constructive and destructive interference, respectively~\cite{HERA}.

As of the beginning of Run II, there have been no dedicated searches for \itev\ extra dimensions. However, a number of constraints have been derived by phenomenological analysis of the existing data. For a review of indirect cosntraints see, e.g. Ref.~\cite{KCGL2}. The best limits come from LEP electroweak precision measurements; the combined limit on the compactification scale of the \itev\ dimensions $M_C \equiv 1/R$ approaches 6.8 TeV.

\begin{table}[hbt]
\caption{95\% C.L. lower limits (in TeV) on the compactification scale $M_C \equiv 1/R$
of the \itev\ ED. From Ref.~\protect\cite{KCGL2}.\label{table:itev} }
\medskip 
\begin{tabular}{lc}
\hline
Experiment and channel      & $M_C^{95}$ (TeV) \\
\hline
LEP~2:                    \\
{} Hadronic cross section, angular distributions, $R_{b,c}$ & 5.3 \\
{} $\mu,\tau$ cross section, angular distributions & 2.8 \\
{} $ee$ cross section, angular distributions & 4.5\\
{} LEP combined          & 6.6 \\
\hline
HERA:     \\
{} $ep \to e + X$ & 1.4 \\
{} $ep \to \nu + X$ & 1.2 \\
{} HERA combined & 1.6 \\
\hline
TEVATRON: \\
{} Drell-Yan           & 1.3 \\       
{} Dijet production      & 1.8 \\
{} Top production & 0.60 \\
{} Tevatron combined   & 2.3 \\
\hline
All combined & 6.8 \\
\hline
\end{tabular}
\end{table}

Finally, as of the beginning of Run II there have been virtually no experimental constraints on Randal-Sundrum extra dimensions or little Higgs models. The only existing ones come from precision electroweak observables and are rather weak.

\section{RECENT TEVATRON SEARCHES FOR EXTRA DIMENSIONS}

Past year has been extremely successful for the Tevatron collider and the CDF and D\O\ experiments. After lagging behind for a while, not only the instantaneous luminosity has been brought up to the design level, but actually exceeded it. A record luminosity of $1.0 \times 10^{32}$ has been achieved in the most luminous store to date, which happened on July 16, 2004. Both the CDF and D\O\ Collaborations have logged nearly 0.5 fb$^{-1}$ of physics data since the beginning of Run II in 2001, of which $\approx 60\%$ came last year. Both experiments function very well, with the average data-logging efficiency of 85\% over the last year. Most of the analyses presented in this paper are based on approximately half of the total recorded data set. The next update based on $\sim 0.5$~fb$^{-1}$ of data is expected in early 2005.

\subsection{D\O\ Search for Large Extra Dimensions in the Monojet Channel}

The monojet channel offers the best direct reach for the fundamental Planck scale at the Tevatron. As was demonstrated in Run I, its sensitivity is superior to that in the monophoton channel. Nevertheless, it is a challenging channel experimentally, since it is plagued by copius instrumental backgrounds, which have to be understood and suppressed.\footnote{As an evidence of experimental challenge, it's worth noting that it took both the CDF and D\O\ Collaborations some four years to complete the monojet analyses in Run I.} Apart from instrumental backgrounds, there is irreducible physics background from the $Z(\nu\nu)+$jet production that forces to require stringent cutoffs on jet transverse energy ($E_T$) and missing transverse energy (\METs) (the latter being a characteristic signature of the graviton escaping in the bulk space).

\begin{figure}
\includegraphics[width=4in]{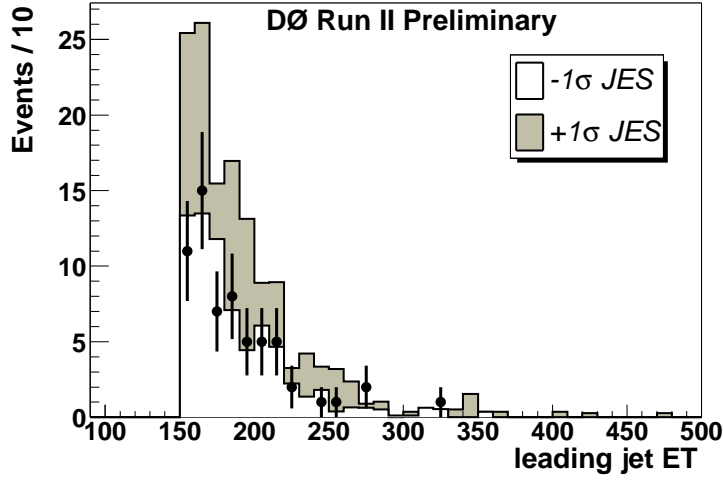}
\vspace{-0.2in}
\caption{Transverse energy of the leading jet in the D\O\ Run II monojet analysis. Points are data, histogram is the expected background, and the shadow band corresponds to the change in the JES of $\pm$ one standard deviation.}
\label{fig:monojet1}
\end{figure}

Recently D\O\ has announced the first preliminary result of searches for large ED in the monojet channel based on new data collected in Run II of the Tevatron. This analysis corresponds to an integrated luminosity of 85 pb$^{-1}$ collected with a special hardware trigger that required the vector sum of transverse energies of jets to exceed 20 GeV. The analysis is based on a subset of the available data collected with this trigger and should be really considered as a proof-of-principle or a warm-up exercise for a more sensitive analysis based on large statistics. Nevertheless, even with this small subset of data an impressive sensitivity has been achieved. Offline, both jet $E_T$ and \MET are required to exceed 150 GeV and a number of quality cuts are introduced to reduce backgrounds. No additional jets with $E_T > 50$ GeV are allowed in the event, and the direction of \MET is required to be at least $30^\circ$ away in azimuth from any jet in the event. That leaves 63 candidates (from the initial sample of 360K events) and corresponds to typical signal efficiency $\sim 5\%$, mainly due to the high $E_T$ and \MET requirements. The number of candidate events agrees within the errors with the expected background of $100 \pm 9 ^{\displaystyle +50}_{\displaystyle -30}$ events, where the second uncertainty is due to the dominant systematics related to the jet energy scale (JES). The background is dominated by the irreducible $Z(\nu\nu) +$jets beackground, with the second most important contribution coming from the $W(\tau\nu) + X$ production where $\tau$ is either missed or reconstructed as a jet. The $E_T$ spectrum of the leading jet is shown in Fig.~\ref{fig:monojet1}.

\begin{figure}
\includegraphics[width=4in]{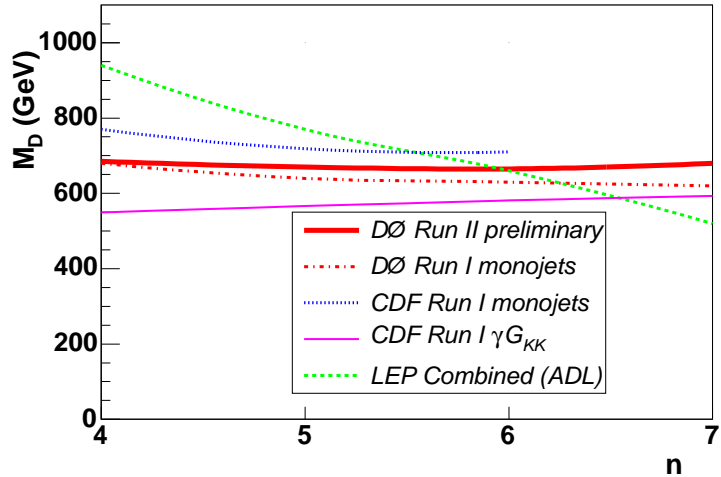}
\vspace{-0.2in}
\caption{95\% CL lower limits on the fundamental Planck scale $M_D$ from the D\O\ monojet analysis in Run II. 
Also shown are the Tevatron Run I ($K = 1$) and LEP limits.}
\label{fig:monojet2}
\end{figure}

Since the data agree with the background prediction, limits on large ED are set. The signal is simulated via PYTHIA~\cite{PYTHIA} Monte Carlo with the implementation of large ED effects of Ref.~\cite{LM}, followed by a full simulation of the D\O\ detector. Uncertainty on the JES is the dominant systematics for the signal. Limits on the fundamental Planck scale $M_D$ from this analysis (for the $K$-factor of 1) are listed in Table~\ref{table:monojet} and shown in Fig.~\ref{fig:monojet2}, together with the Tevatron Run I and LEP constraints. Note that they already improve on the D\O\ Run I limits, although are still shy of the CDF Run I limits. It's worth mentioning that the Tevatron and LEP searches are complementary: Tevatron has higher sensitivity for high number of ED $n$, while LEP offers better limits for small $n$.

\begin{table}[hbt]%[H] add [H] placement to break table across pages
\caption{\label{table:monojet}
95\% CL lower limits on the fundamental Planck scale $M_D$ (in TeV) in the ADD model from the D\O\ Run II analysis.}
\begin{tabular}{cccc}
% Lines of table here ending with \\
\hline
 $n=4$ & $n=5$ & $n=6$ & $n=7$\\
\hline
 0.68  & 0.67  & 0.66  & 0.68 \\
\hline
\end{tabular}
\end{table}

Since the analysis has been completed, the JES systematics has been reduced by a factor of two. Together with much higher statistics recorded with the new trigger, the next update of the D\O\ analysis in early 2005 is expected to reach new level of sensitivity to large ED and supersede previous Tevatron and LEP results.

\subsection{Searches for Large Extra Dimensions via Virtual Graviton Effects}

Both Tevatron collaboration pursue search for large ED in Drell-Yan and diphoton production in Run II. Recent CDF analysis is based on 200 pb$^{-1}$ of dielectron data. Both electrons require to have large transverse energy and satisfy standard quality criteria. At least one of the electrons is required to be in the central region of the detector (i.e., $|\eta| < 1.0$, where $\eta$ is pseudorapidity measured relative to the geometrical center of the detector). The second electron can be either central or forward ($1.1 < |\eta| < 2.8$).

\begin{figure}
\includegraphics[width=3.1in]{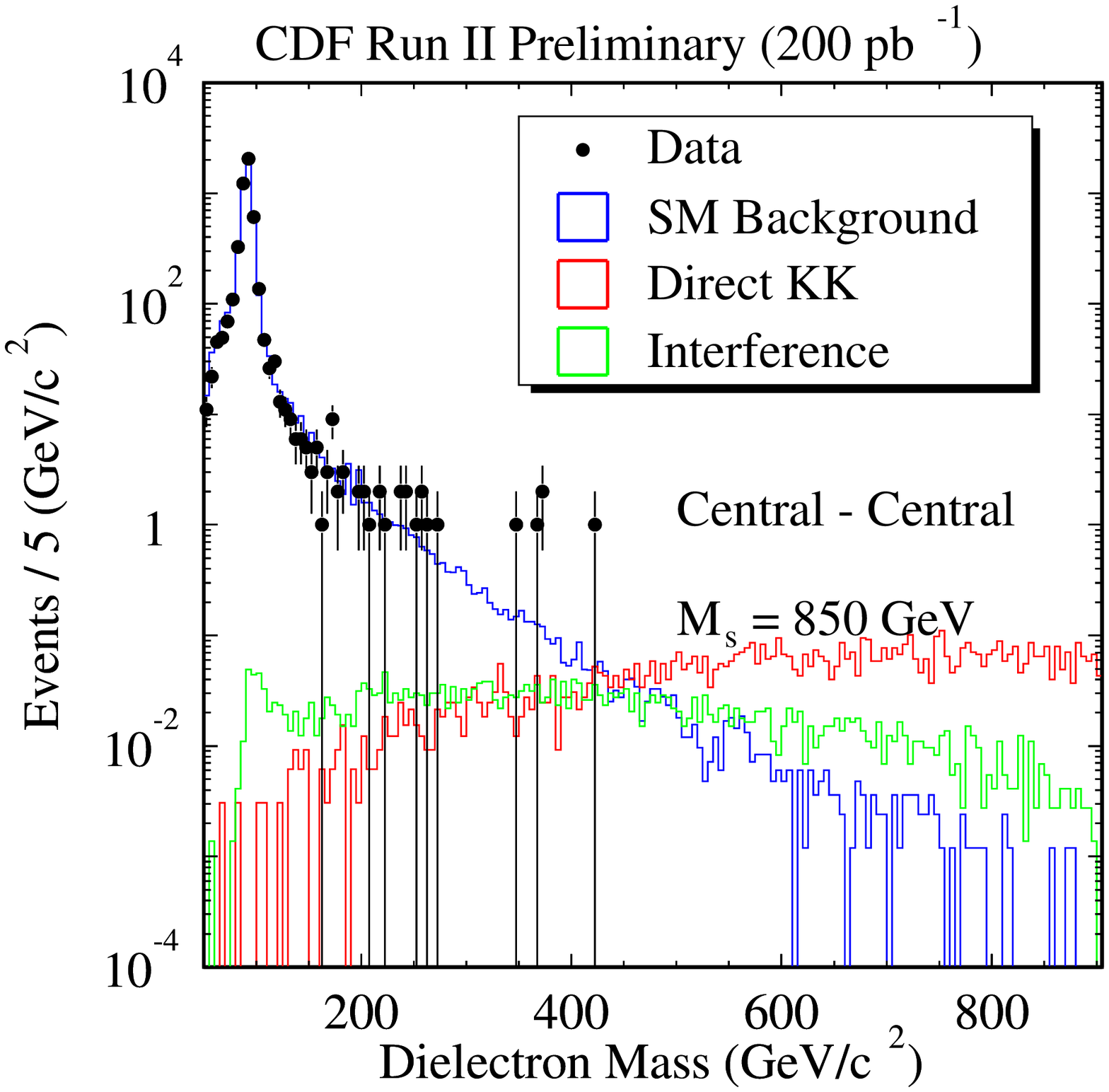}
\includegraphics[width=3.1in]{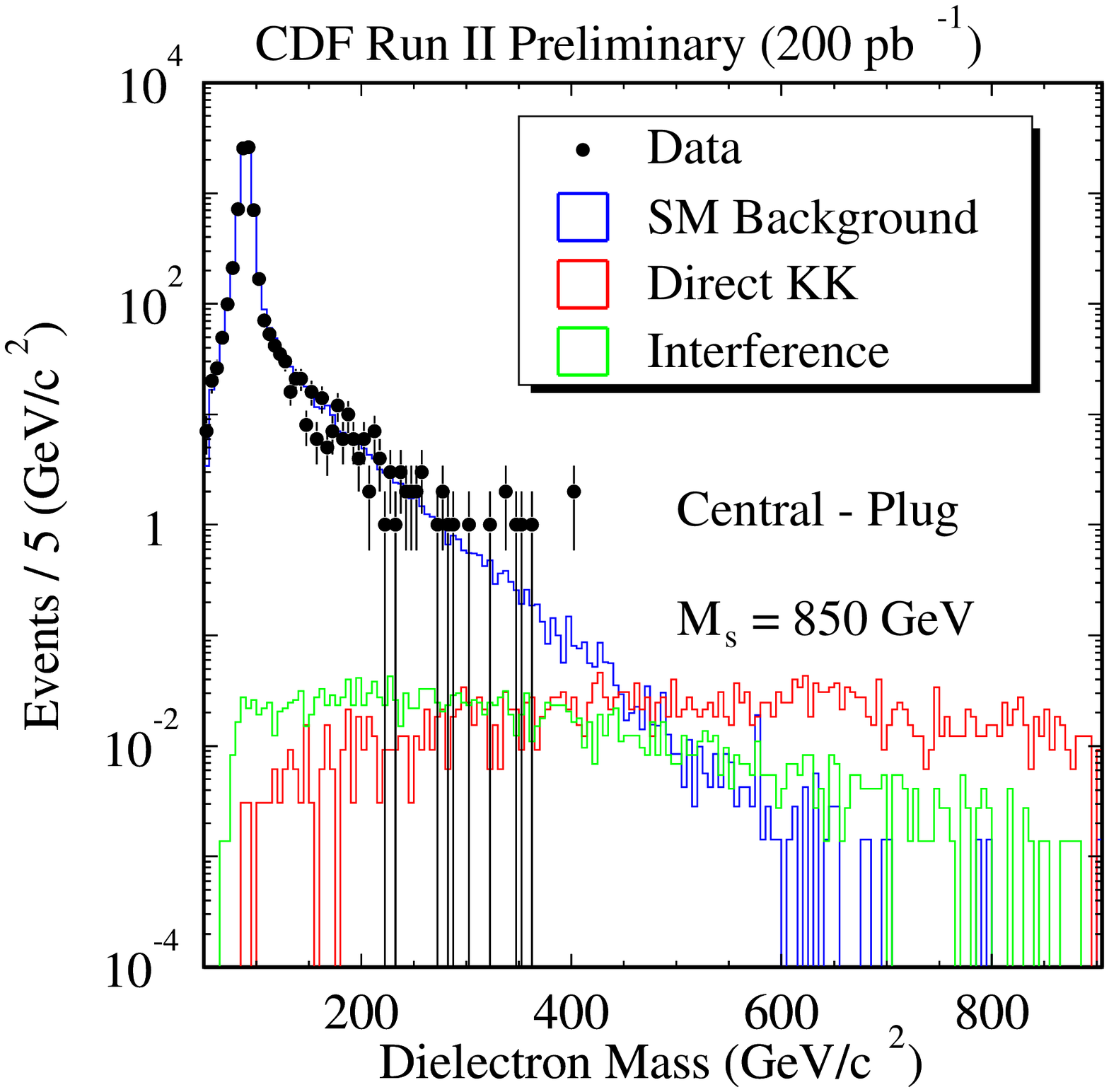}
\caption{Dielectron mass spectrum from the CDF search for large extra dimensions via virtual graviton effects. Left: central-central; right: central-forward combinations. Points are data; solid line is the sum of SM backgrounds. Also shown in the plots are effects of virtual graviton exchange due to the interference and direct terms for $M_S = 850$ GeV in Hewett's convention.}
\label{fig:CDFee}
\end{figure}

The invariant mass spectra of the dielectrons for two different topologies are shown in Fig.~\ref{fig:CDFee}. They agree well with the expected background dominanted by the DY production. The shape of the contributions from virtual graviton exchange for $M_S = 850$~GeV (in Hewett's convention) are also shown in the figure. As seen from the plot, the dominant contribution in the region where signal exceeds expected background is from the direct gravity term (see Eq. (\ref{eq:virtual})), rather than the interference term. Given no evidence for large ED, limits on the ultraviolet cutoff $M_S$ are set within different formalisms and listed in Table~\ref{table:virtual}. A LO event generator of Ref.~\cite{Baur} has been used to generate signal, with NLO effects on the signal accounted for via a constant $K$-factor of 1.3, similar to that for DY~-- the choice which has been recently justified by exact NLO calculations of virtual graviton production and decay into a lepton pair~\cite{K-factor}.

\begin{table}[htb]
\begin{tabular}{lc@{}cccccc@{}cc}
\hline
Experiment & GRW~\cite{GRW} & \multicolumn{6}{@{}c}{HLZ~\cite{HLZ}} & \multicolumn{2}{@{}c}{Hewett~\cite{Hewett}} \\
\hline
& & ~~$n$=2 & $n$=3 & $n$=4 & $n$=5 & $n$=6 & $n$=7~~ & ~~$\lambda=+1$ & $\lambda = -1$\\
CDF Run II          & 1.06 &      & 1.32 & 1.11 & 1.00 & 0.93 & 0.88 & 0.99 & 0.96  \\
\hline
D\O\ Run II         & 1.36 & 1.56 & 1.61 & 1.36 & 1.23 & 1.14 & 1.08 & 1.22 & 1.10  \\
\hline
D\O\ Run I + Run II & 1.43 & 1.67 & 1.70 & 1.43 & 1.29 & 1.20 & 1.14 & 1.28 & 1.14  \\
\hline
\end{tabular}
\caption{95\% CL lower limits on the ultraviolet cutoff $M_S$ (in TeV) from the Tevatron Run II, within several phenomenological frameworks. NLO QCD effects have been accounted for via a $K$-factor of 1.3.}
\label{table:virtual}
\end{table}

The D\O\ analysis is also based on 200 pb$^{-1}$ of data and utilizes both the diphoton and dielectron channels. In order to maximize the selection efficiency tracking information is not used in the selection. Consequently, electrons are not distinguished from photons, which allows one not to lose the events in which one of the photons has converted or one of the electrons does not have a reconstructed track. In what follows, such electrons or photons reconstructed in the calorimeter are referred to as ``EM objects.'' Like in the CDF case, one of the EM objects is required to be central ($|\eta| < 1.1$), while the other one can be either central or forward ($1.5 < |\eta| < 2.4$). 

Another important technique used by D\O\ is the two-dimensional fits in the $(M_{\rm EM-EM},|\cos\theta^*|)$-plane, where $\theta^*$ is the scattering angle in the c.o.m. frame. These two variables define the $2 \to 2$ pair-production process completely and thus contain full information about the interaction Lagrangian, which allows to achieve maximum sensitivity to gravity effects.

Backgrounds in this search are from the SM DY and direct diphoton production, as well as from jets misidentified as EM objects. The latter contribution is determined from data by inverting quality cuts on the EM objects.

\begin{figure}
\includegraphics[width=4.5in]{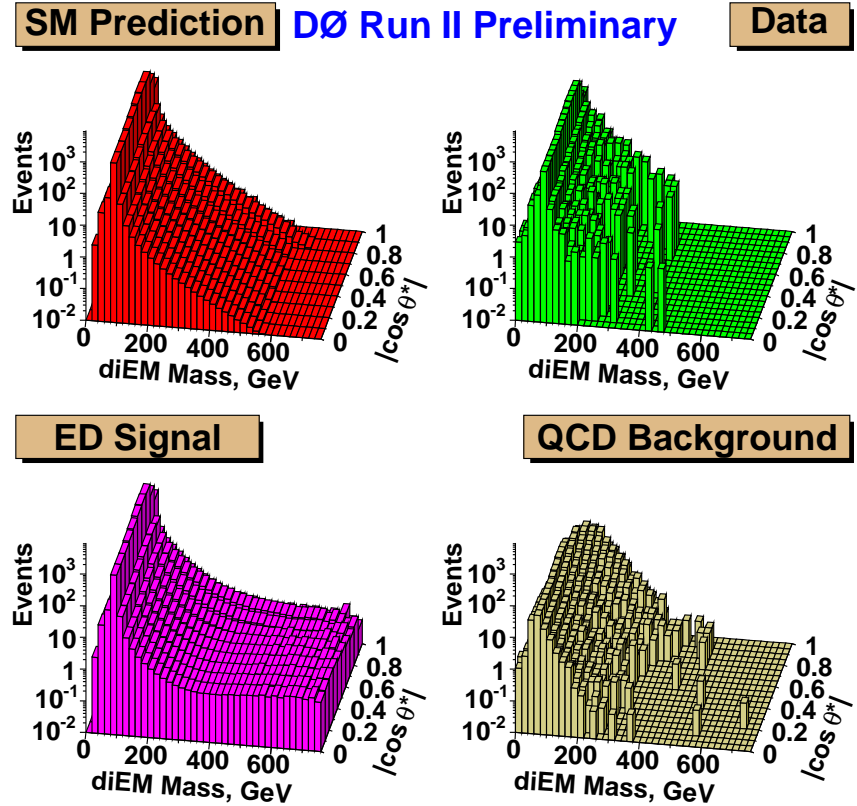}
\vspace{-0.1in}
\caption{
\label{fig:D0EMEM}
The diEM mass vs. $|\ct|$ distribution for the D\O\ diEM candidate sample. Top left: SM background from Drell-Yan and direct diphotons; top right: data; bottom left: ED signal plus the SM background for $\eta_G = 2.0$; bottom right: instrumental background.}
\end{figure}

The comparison of the D\O\ data with the expected SM and instrumental backgrounds, as well as the ED effects are shown in Fig.~\ref{fig:D0EMEM}. Note that the SM tail at high values of $|\cos\theta^*|$ and high masses (due to a collinear divergence in the diphoton production), which would result in decreased sensitivity to large ED effects if only the mass variable is used, does not affect the sensitivity of the 2D-method, as the effects of virtual graviton exchange are pronounced mostly at high masses and low values of the $|\cos\theta^*|$.

The signal and SM backgrounds are modeled via a LO Monte Carlo generator of Ref.~\cite{KCGL1}, augmented with a fast parameteric simulation of the D\O\ detector, which takes into account the $p_T$-boost of the diEM system due to initial state radiation, detector resolution, and other instrumental effects. NLO effects for both signal and SM backgrounds are accounted for via a constant $K$-factor of 1.3.

Since the data agree with the sum of instrumental and SM backgrounds, limits on $M_S$ are set using the template method via a fit to a 2D-extension of Eq.~(\ref{eq:virtual}). The SM, interference, and direct terms are calculated as 2D-distributions in the $(M_{\rm EM-EM},|\cos\theta^*|)$-plane and the data is fit to Eq. (\ref{eq:virtual}), with $\eta_G$ being a free parameter of the fit. The limits obtained as a result of this fit are listed in Table~\ref{table:virtual}. These are the most stringent limits on large ED to date, which can be further improved by combiing them with the D\O\ Run I limits~\cite{D0-diem}, as detailed in Table~\ref{table:virtual}. Note that for $n=2$ these limits are very similar to the sensitivity obtained in the tabletop gravity measurement experiments.

\begin{figure}[tbhp]
\includegraphics[width=3.3in]{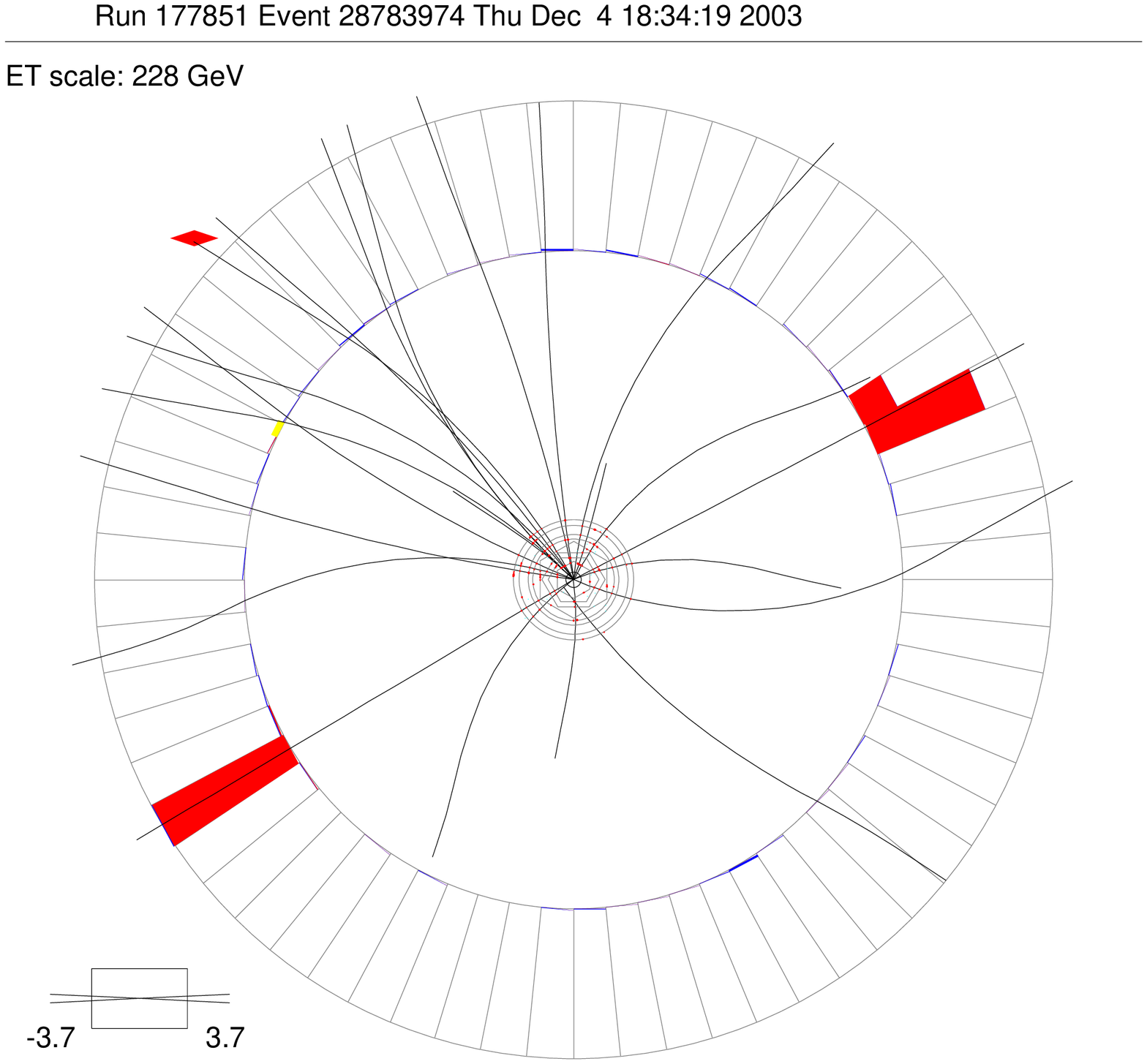}
\vspace{0.1in}
\includegraphics[width=3.3in]{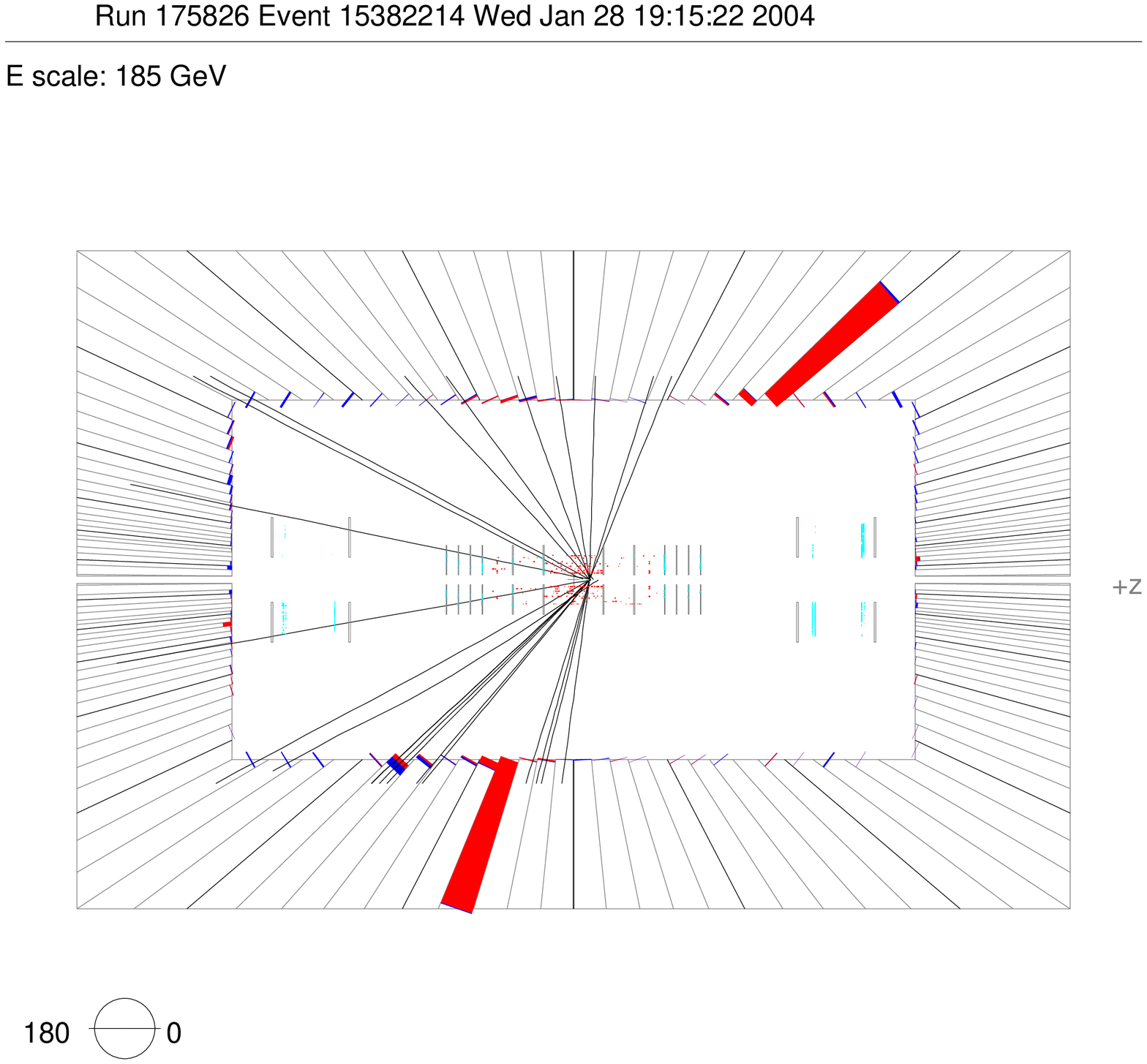}
\caption{
\def\baselinestretch{1}
Left pane: event display of the highest-mass D\O\ $e^+e^-$ candidate with the invariant mass of 475 GeV and $|\ct|$ of 0.01 (Run 177851, Event 28783974, or ``Event Callas''). Right pane: event display of the highest-mass D\O\ $\gamma\gamma$ candidate with the invariant mass of 436 GeV and $|\ct|$ of 0.03 (Run 175826, Event 15382214, or ``Event Farrar''). This is the highest mass diphoton event ever observed.  While in agreement with SM background prediction, both candidates have kinematical characteristics that would be typical of a signal from large ED.}
\label{fig:D0events}
\end{figure}

While the data agree well with the SM predictions, the two highest-mass candidate events observed in the analysis are quite interesting. Their parameters are listed in Table~\ref{table:events} and their displays are shown in Fig.~\ref{fig:D0events}. One of them is an $e^+e^-$ event, while the other one is a diphoton. Both have anomalously low value of $|\cos\theta^*|$, which make them typical of the topologies expected from the virtual graviton exchange. Nevertheless, there is no evidence for any excess above the SM predictions in the D\O\ analysis. More data would show whether there is any significance in observing such low $|\cos\theta^*|$ for the highest-mass candidate events.

\begin{table}[htbp]
\begin{tabular}{|c|c|c|c|c|c|c|c|c|c|c|c|c|}
\hline
Run     & Event & \MET & Type       & $E_T^1~(p_T^1)$ & $E_T^2~(p_T^2)$& $\eta_1$ & $\eta_2$ & $\phi_1$ & $\phi_2$ & $M$     & \ct   & $N_{\rm jet}$ \\ 
\hline
177851 & 28783974 	& 8.8 & $e^+/e^-$ & 239.2~(190.8)   & 226.9~(234.4) & -0.45 & -0.50 & 0.48 & 3.68 & 475 & 0.01 & 1 \\
175826 & 15382214 & 17.0 & $\gamma\gamma$ &  197.9   &  158.4  & -0.47   & 0.87  & 3.20 & 0.06 & 436 & 0.03 & 0\\
\hline
\end{tabular}
\caption{Parameters of the two D\O\ highest diEM-mass candidates. All energy variables are in GeV.
\label{table:events}}
\end{table}

\subsection{Search for \itev\ Extra Dimensions via Virtual Effects}

While the sensitivity of the Tevatron to the \itev\ ED is expected to be inferior to that from precision measurements at LEP, a much higher energy of the Tevatron would make  the results complementary to those at LEP. Since there has been no dedicated experimental searches for \itev\ ED done at high energies, it's worth performing such a search at the Tevatron. The D\O\ experiment has recently announced the first preliminary result of a search for \itev\ ED

The analysis follows closely the search for large ED via virtual effects, with the exception that in the case of \itev\ ED, the modification of the SM cross section is due to an exchange of virtual Kaluza-Klein excitations of the SM gauge bosons, and not the graviton. In the case of the D\O\ analysis, the effects of the KK towers of $\gamma$ and $Z$ have been sought. Since neither $\gamma$ nor $Z$ couples to photons in the SM, only the dielectron channel has been used for the analysis. The dielectrons were selected from the 200 \ipb diEM sample used in the large ED analysis by requiring at least one track with a good spatial match with one of the EM objects. The analysis has been performed in the same 2D-plane as the large ED analysis, as has been suggested in Ref.~\cite{KCGL1}.

The effects of the exchange of the KK towers of the $\gamma/Z$ can be parameterized via a bilinear function similar to that of Eq. (\ref{eq:virtual}), where the parameter $\eta_G$ is replaced with $\eta_C=\pi^2/(3M_C^2)$. Here $M_C$ is the compactification scale, $M_C = 1/R$.

\begin{figure}
\includegraphics[width=4.5in]{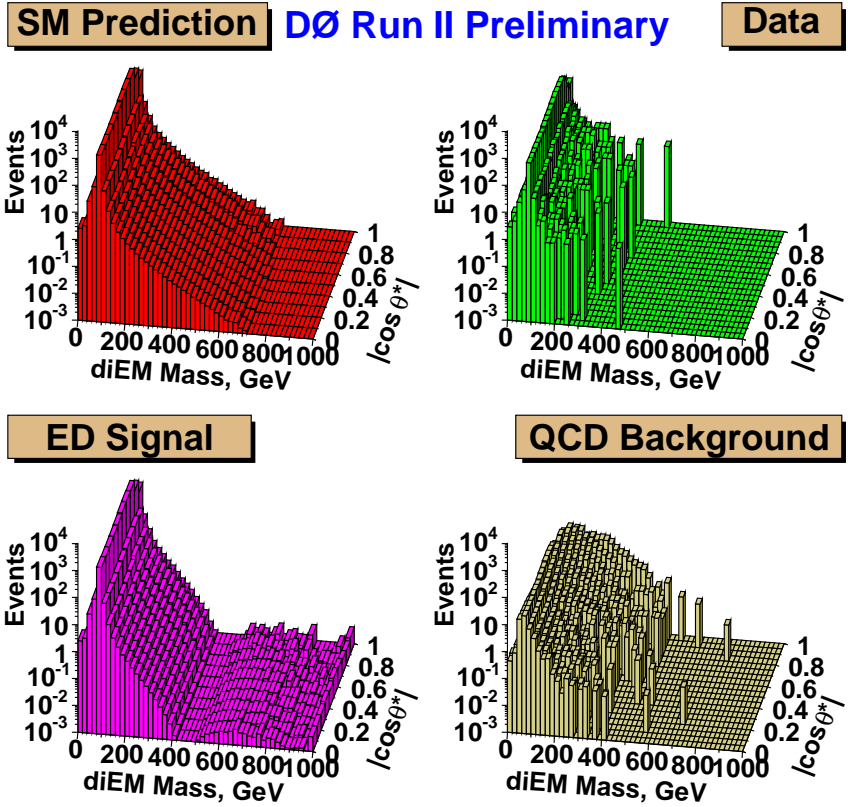}
\vspace{-0.1in}
\caption{
\label{fig:itev}
The $e^+ e^-$ mass vs. $|\ct|$ distribution in the D\O\ \itev\ analysis.
Top left: SM Drell-Yan background; top right: data; bottom left: \itev ED signal plus the 
SM background for $\eta_C = 5.0$~TeV$^{-2}$ (or $M_C \approx 0.8$~TeV); bottom right: instrumental background.}
\end{figure}

The data is compared to the SM background, instrumental background, and the effects from \itev\ ED in Fig.~\ref{fig:itev}. As in the large ED analysis, a good agreement is observed between the data and the sum of backgrounds. The highest-mass event in the plot is the Event Callas from the large ED analysis. 

Note that the effects of the $\gamma_{\rm KK}/Z_{\rm KK}$ exchange are more convoluted than those of the $G_{\rm KK}$ exchange. A deficit of events is observed at masses $\sim M_C/2$, which corresponds to the negative interference between the DY and the first KK modes of $\gamma$ and $Z$. An enhancement of the cross section is observed at masses $\sim M_C$ due to resonant production of the first KK modes of $\gamma$ and $Z$ with low degree of virtuality.

The data is fit to the sum of the SM, interference, and direct Kaluza-Klein terms with $\eta_C$ being a free parameter. The following 95\% CL lower limit on $M_C$ has been obtained:
\begin{equation}
	M_C > 1.12~\mbox{TeV},\label{eq:itev}
\end{equation}
which corresponds to the limit on the size ot \itev\ ED $R < 1.75\times 10^{-19}$~m.

\subsection{Searches for Randall-Sundrum Gravitons in the Dilepton and Diphoton Channels}

\begin{figure}
\includegraphics[width=4in]{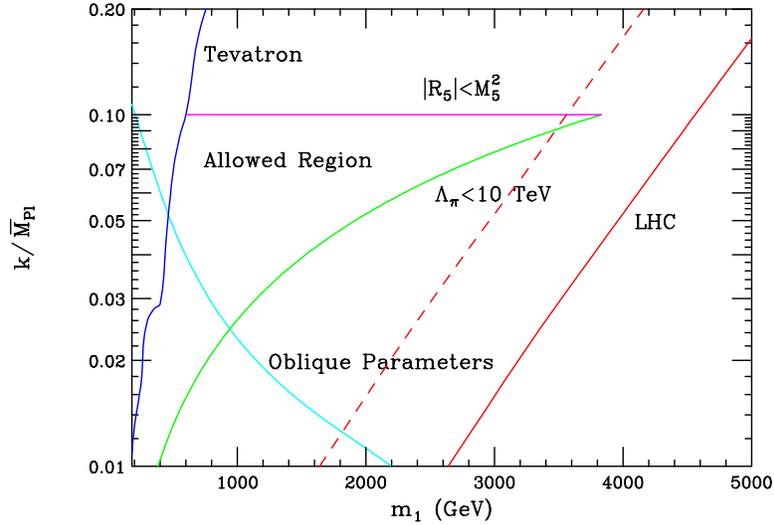}
\vspace{-0.1in}
\caption{
\label{fig:RS_allowed}
Allowed range of the parameters $M_1$ (called $m_1$ in the plot) and $k/\overline{M}_{\rm Pl}$ in the Randall-Sundrum model. The allowed region is the tooth-like shape restricted by the limits from the oblique parameters and theoretical bounds coming from the requirement of $\Lambda_\pi < 10$~TeV and  the limit on the curvature of the AdS space (horizontal line). (Both are required in order not to create another large hierarchy of scales.) Also shown in the plot are the expected sensitivity in the DY channel at the Tevatron with 2 fb$^{-1}$ of data and at the LHC with integrated luminosities of 10 fb$^{-1}$ (dashed line) and 100 fb$^{-1}$ (solid line marked ``LHC''). From Ref.~\protect\cite{RSext}.
}
\end{figure}

Both the CDF and D\O\ collaborations performed first searches for the effects of Kaluza-Klein gravitons in the Randall-Sundrum model. As discussed above, the simplest RS model is fixed by specifying just two parameters: the curvature of the AdS space $k$ and the radius of compactification $R$. However, from the experimental point of view, it is more convenient to work with an equivalent set of parameters, which correspond to direct observables: $M_1$, the mass of the first Kaluza-Klein excitation of the graviton, and a dimensionless parameter $k/\overline{M}_{\rm Pl}$, which governs the coupling of the gravitons to the SM fields and hence defines the internal width of the KK gravitons. The allowed range for these two parameters is shown in Fig.~\ref{fig:RS_allowed}~\cite{RSext}.

Both the CDF and D\O\ Collaborations used the dilepton and diphoton mass spectra and searched for a narrow resonance, which would indicate a production of the first excitation of the KK graviton. Note that even for the largest allowed value of $k/\overline{M}_{\rm Pl} \sim 0.1$, the internal width of the first excitation is significantly narrower than the mass resolution of either the CDF or D\O\ detector, so the search has been only optimized as a function of graviton mass. 

\begin{figure}
\includegraphics[width=3.5in]{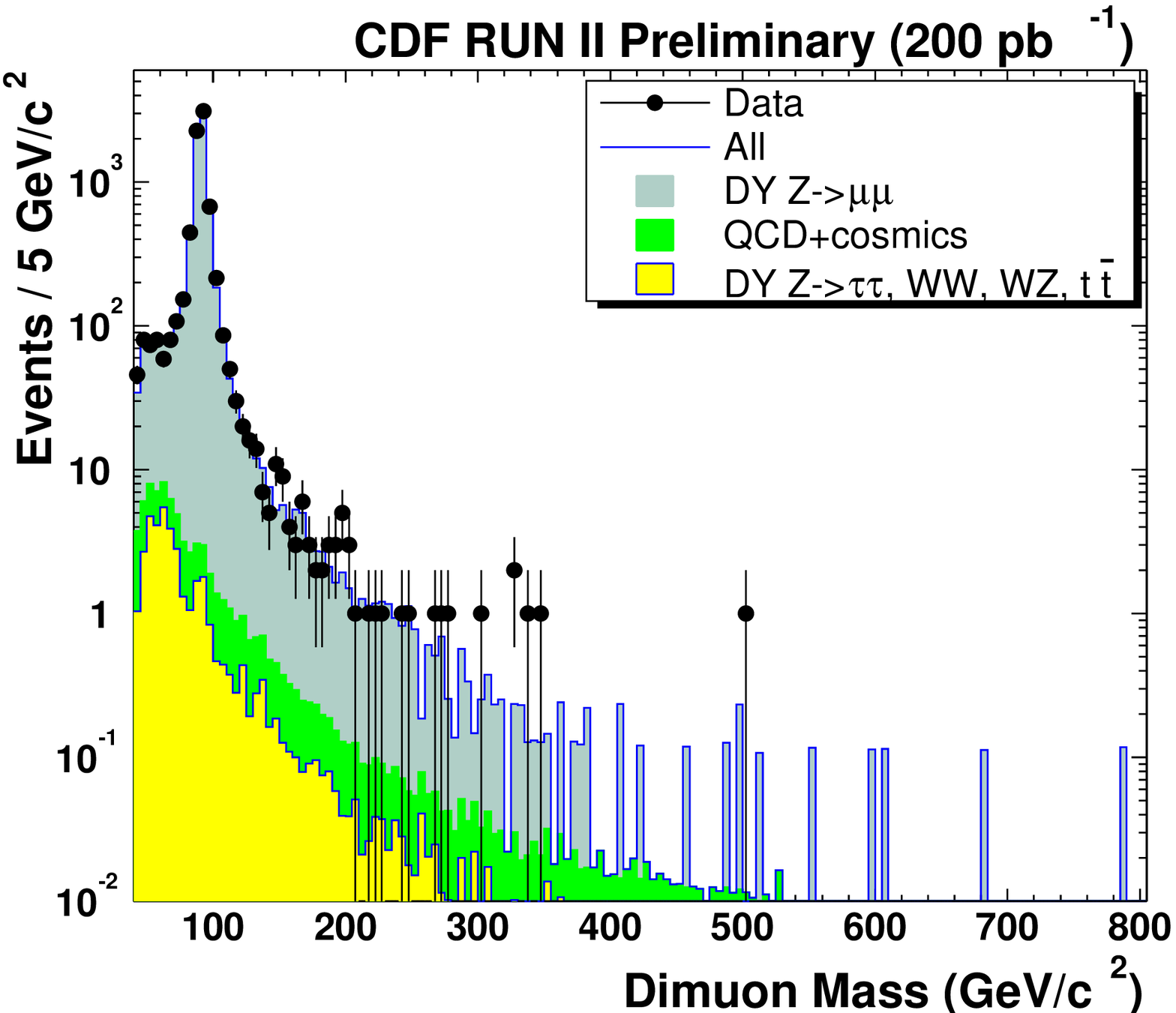}
\includegraphics[width=3.1in]{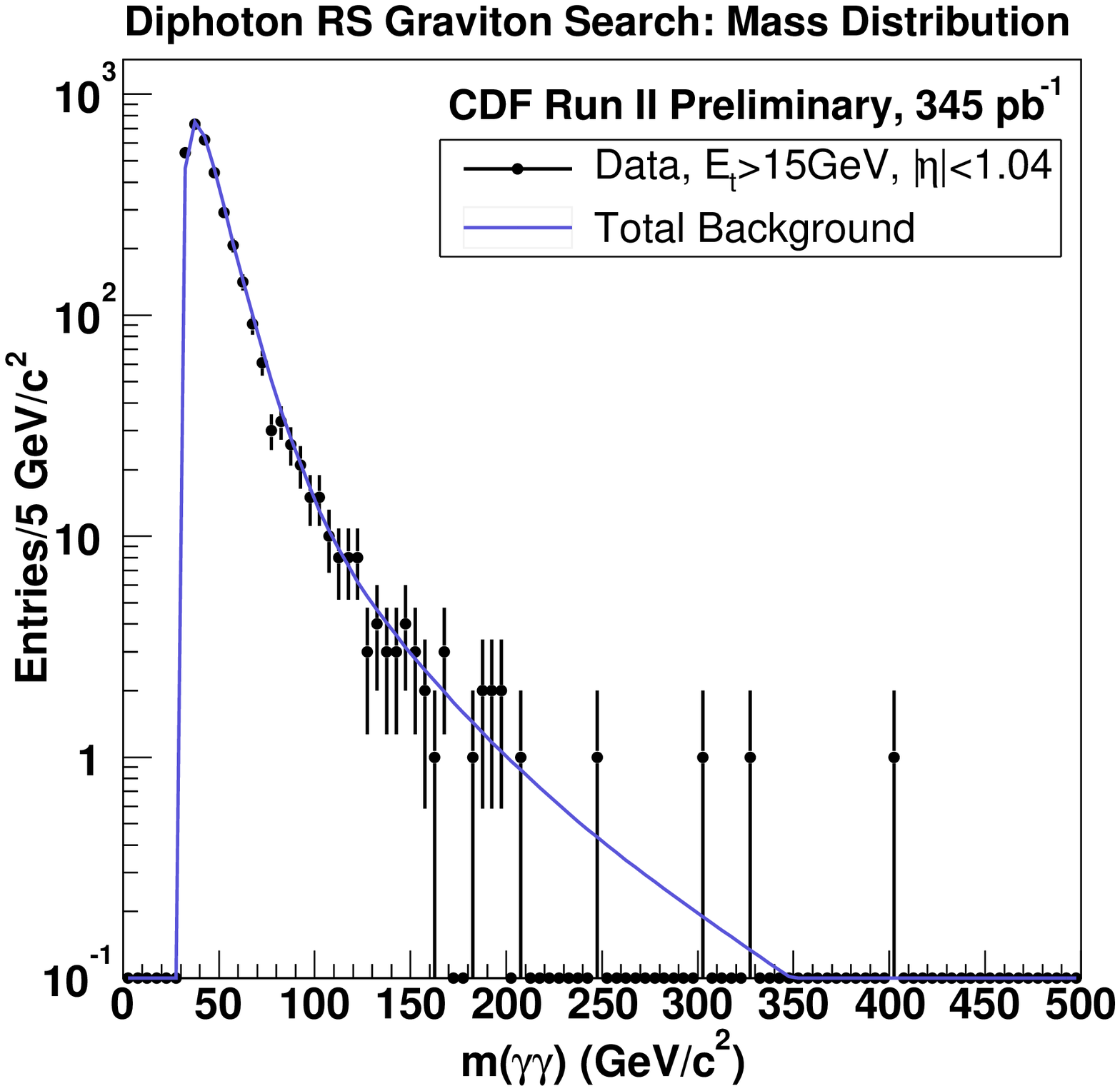}
\caption{
\label{fig:CDFuugg}
Dimuon (left) and diphoton (right) mass spectrum from the CDF search for Randall-Sundrum gravitons. Points are data; solid line is the total expected background.}
\end{figure}

\begin{figure}
\includegraphics[width=3.5in]{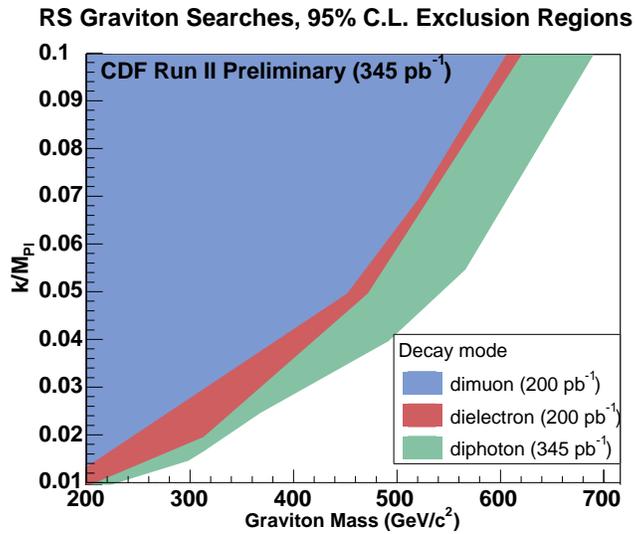}
\vspace{-0.1in}
\caption{\label{fig:CDFRS}
Individual 95\% CL limits on the RS model parameters $M_1$ and $k/\overline{M}_{\rm Pl}$ from the CDF search in three channels. The shaded area to the left of the corresponding curve is excluded.}
\end{figure}

\begin{figure}
\includegraphics[width=3.1in]{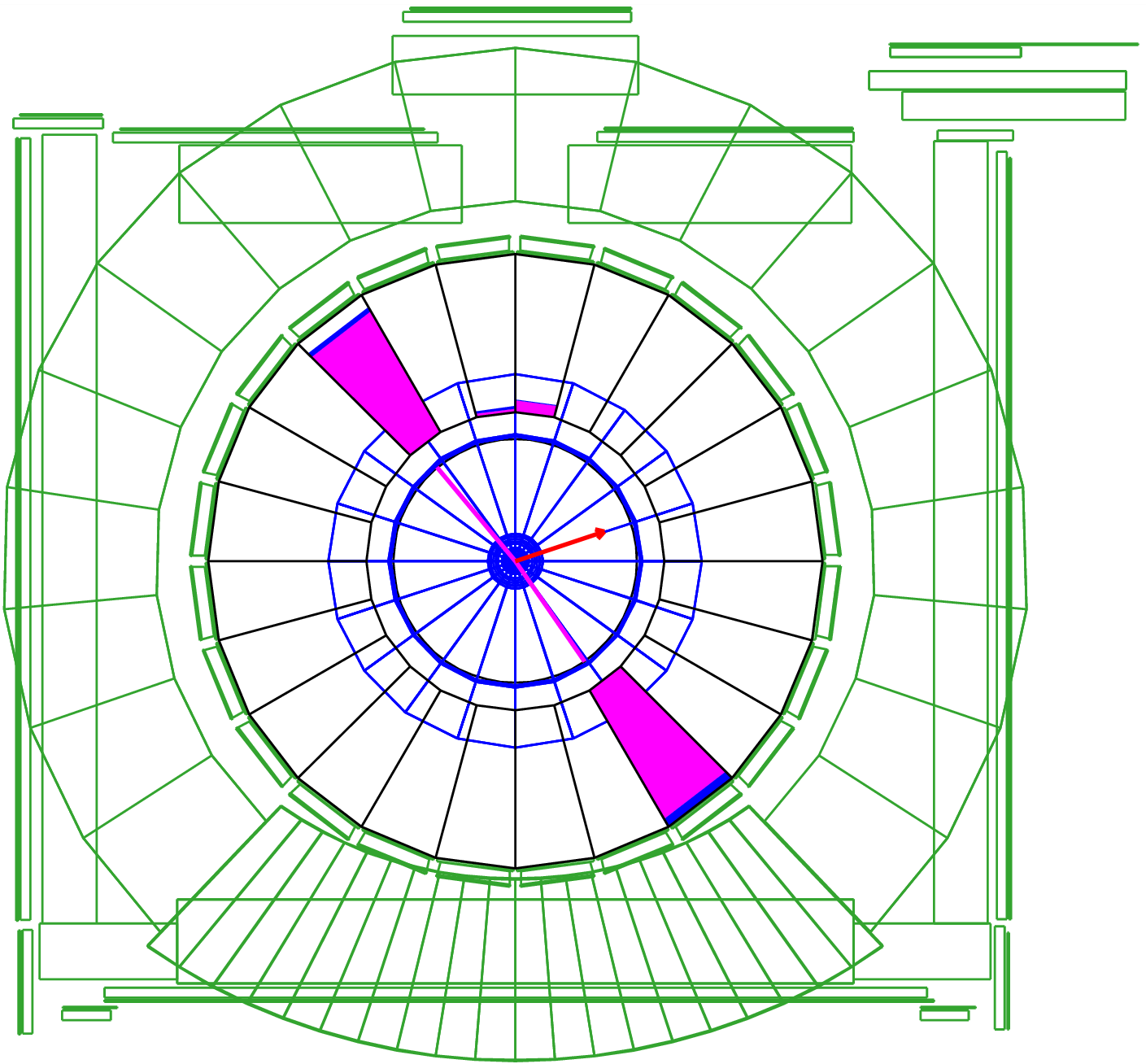}
\includegraphics[width=3.1in]{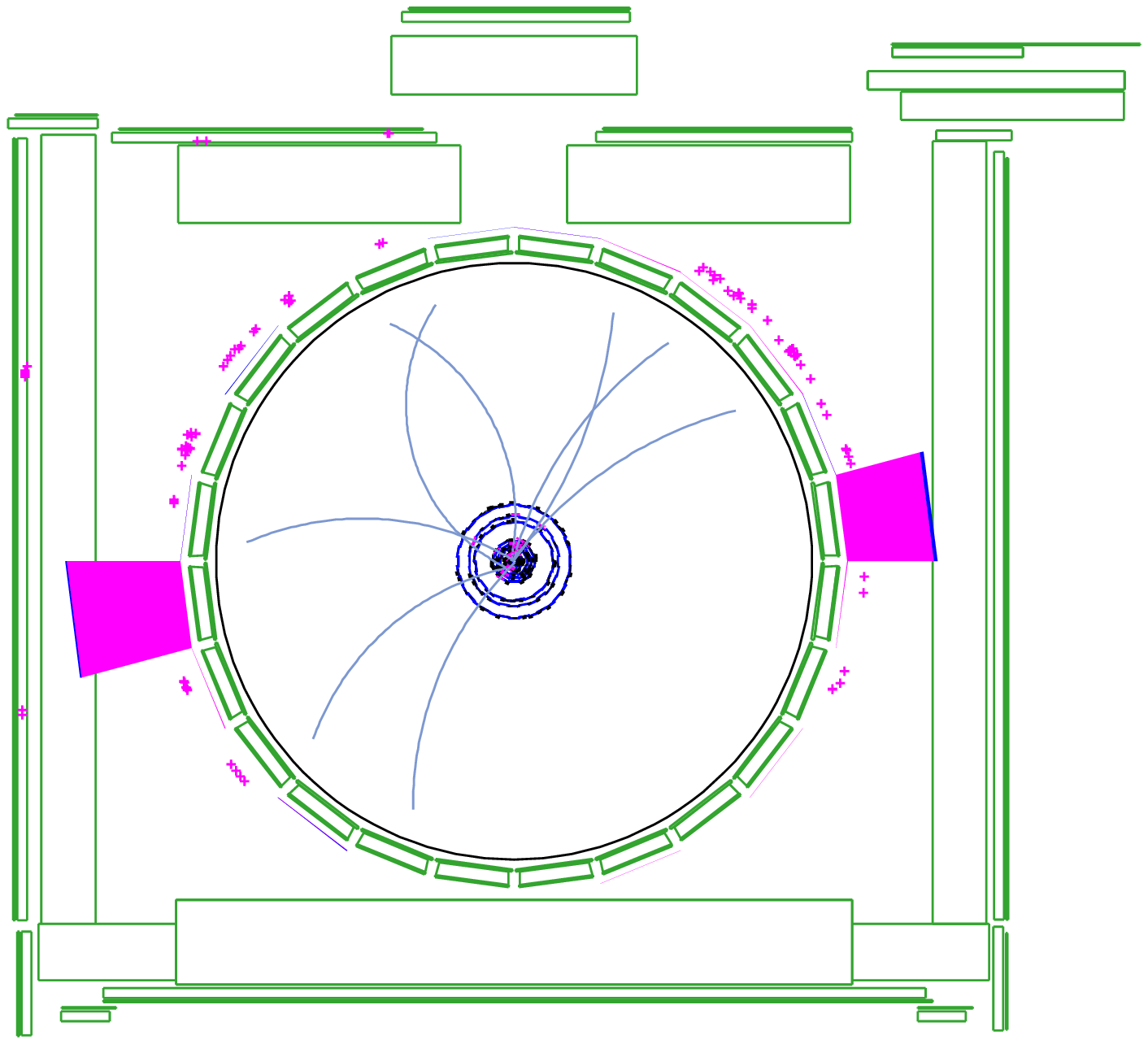}
\vspace{-0.1in}
\caption{
Left: an event display for one of the high-mass ($M_{ee} = 371$ GeV) CDF dielectron candidate events.
Right: an event display for the highest-mass ($M_{\gamma\gamma} = 405$ GeV) CDF diphoton candidate event.}
\label{fig:CDFevent}
\end{figure}

The dielectron mass spectrum used in the CDF analysis is the same as used in the large ED analysis and is shown in Fig.~\ref{fig:CDFee}. The dimuon and diphoton mass spectra are shown in Fig.~\ref{fig:CDFuugg} and correspond to the integrated luminosity of 200 \ipb and 345 \ipb respectively. All three spectra are described well by the background-only hypothesis; no narrow peaks are seen in any of the spectra. Signal cross section is estimated with PYTHIA~\cite{PYTHIA} Monte Carlo, with the NLO effects accounted for by multiplying the cross section by a constant $K$-factor of 1.3. At each hypothetical graviton mass a window with the size of $\pm 3$ mass resolutions is used to count the events and the results of these counting experiments are interpreted as limits on the RS graviton production. These limits are shown in Fig.~\ref{fig:CDFRS} for each of the three channels. The sensitivity is dominated by the diphoton channel, not only due to the higher statistics, but also due to the fact that branching fraction of RS gravitons into diphotons is twice as large as that into a single dilepton channel. Event displays of one of the highest-mass dielectron and the highest-mass diphoton candidates from the CDF search are shown in Fig.~\ref{fig:CDFevent}.

\begin{figure}
\includegraphics[width=3.7in]{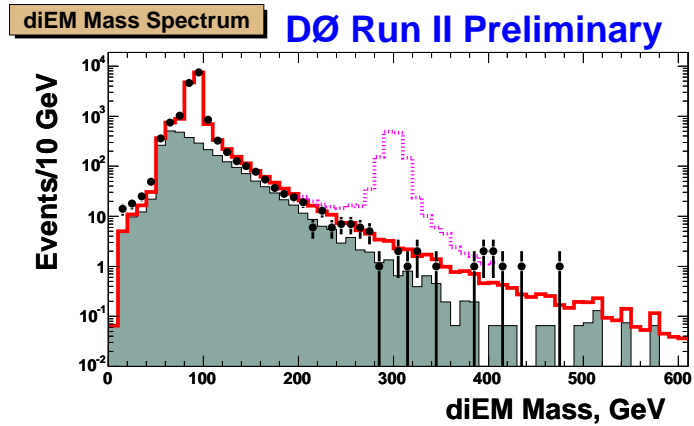}
\vspace{-0.1in}
\caption{
\label{fig:D0mass}
DiEM mass spectrum from the D\O\ search for Randall-Sundrum gravitons. Points are data; the solid line is the total expected background; the shaded histogram shows instrumental background due to jets misidentified as EM objects. Also shown in the plot is a signal for a RS-graviton with the mass of 300 GeV and an arbitrary cross section normalization.}
\end{figure}

\begin{figure}
\includegraphics[width=3.5in]{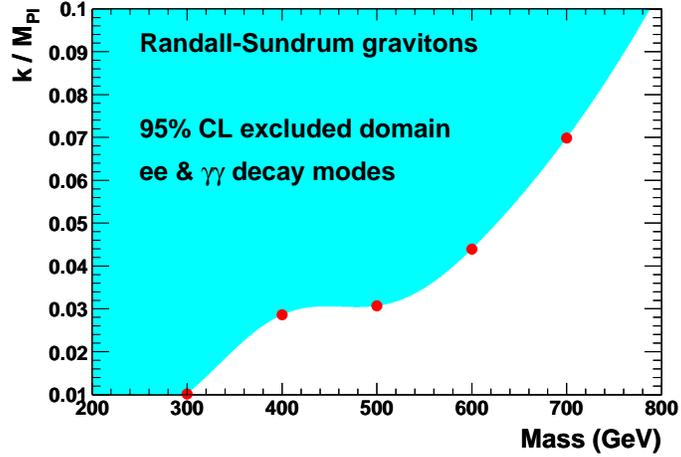}
\caption{\label{fig:D0RS}
Combined 95\% CL limits on the RS model parameters $M_1$ and $k/\overline{M}_{\rm Pl}$ from the D\O\ search in the dielectron and diphoton channels. Dots correspond to the individual points where the counting window optimization has been done. The shaded area is excluded.}
\end{figure}

The corresponding D\O\ search is based on the diEM candidate sample used for the search for large ED. Similar to the latter search, electrons and photons are not distinguished from each other, which allows to combine the two channels without losing any efficiency. The diEM mass spectrum is shown in Fig.~\ref{fig:D0mass}, which is a projection of the 2D-distribution in Fig.~\ref{fig:D0EMEM} on the mass axis. At low masses the mass window for the counting experiment is optimized to yield maximum Gaussian significance $(S/\sqrt{B})$ using the method of Ref.~\cite{KMGL}, modified for exponentially falling background. At high masses, where background is small, the window size is chosen to be $\pm 3$ mass resolutions. Signal is generated with PYTHIA~\cite{PYTHIA} Monte Carlo, followed by a fast detector simulation. The event generator described in the large ED analysis is used to generate SM backgrounds. As in the large ED analysis, the LO cross sections for signal and background is multiplied by a constant $K$-factor of 1.3. As in the CDF analysis, no evidence for narrow resonances is found in the D\O\ data, and the results are presented as limits on the RS model parameters and shown in Fig.~\ref{fig:D0RS}. These limits are the most stringent constraints on the RS model parameters to date. Note that they already exceed theoretical expectations for ten times higher statistics (see \ref{fig:RS_allowed}), due to a thorough optimization of the analysis and an inclusion of the diphoton channel.

\subsection{Search for Extra Gauge Bosons in the Little Higgs Model}

\begin{figure}
\includegraphics[width=2.72in]{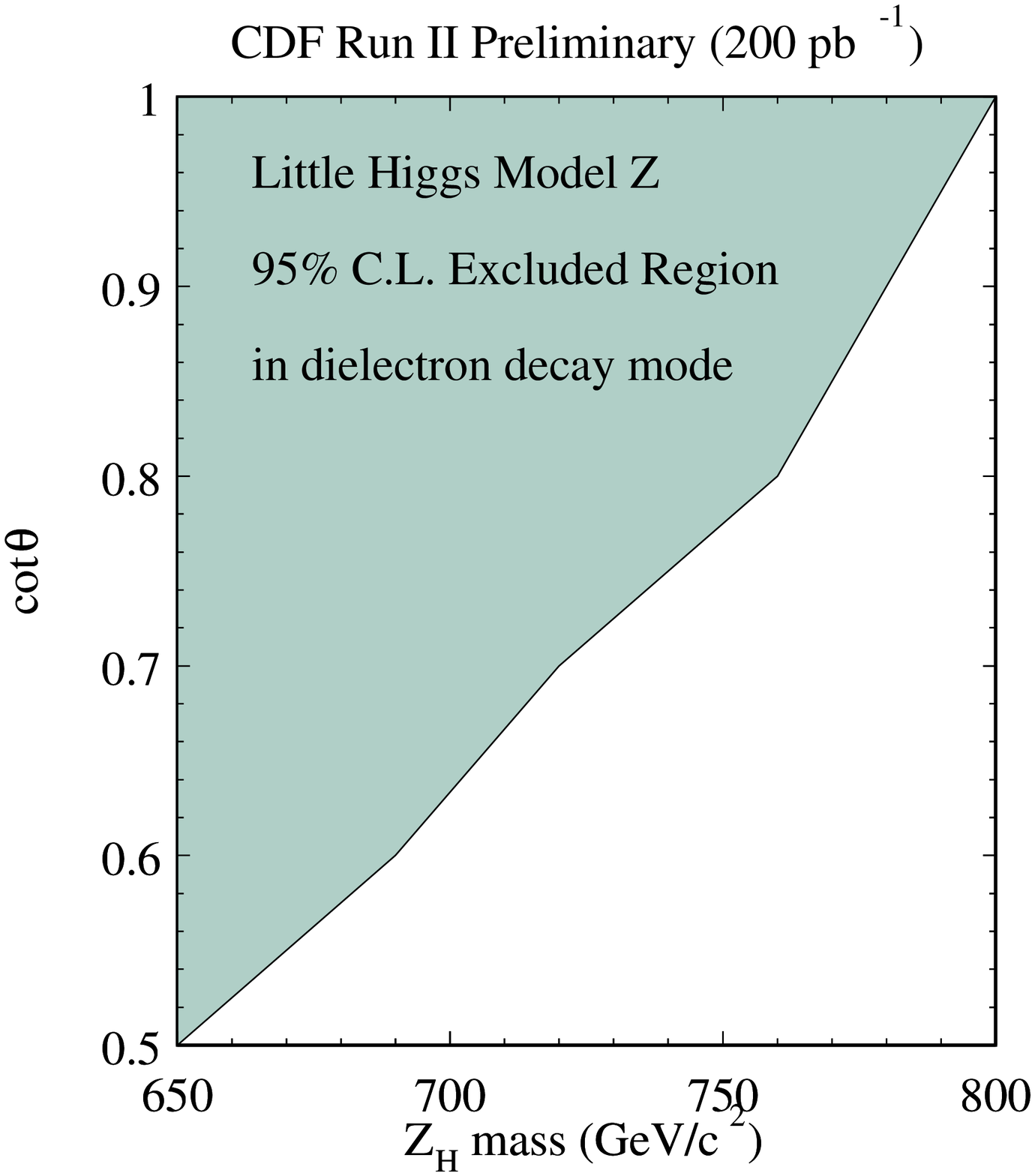}
\includegraphics[width=3.3in]{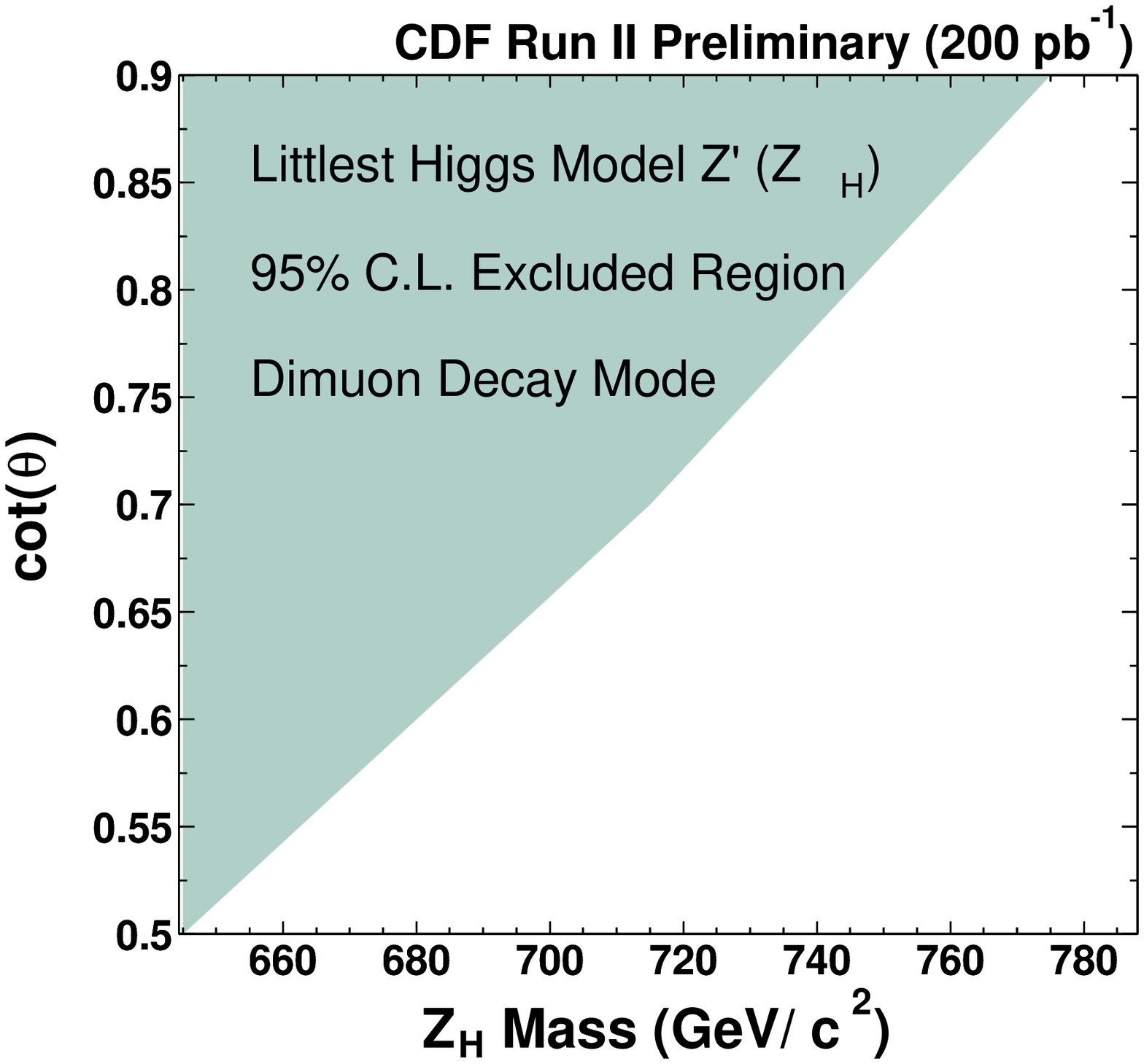}
\caption{\label{fig:LH}
Individual 95\% CL limits on the $Z_H$ mass as a function of the mixing parameter $\cot\theta$ in the little Higgs model of Ref.~\protect\cite{LH}. Left: dielectrons; right: dimuons. In each case the shaded area is excluded.}
\end{figure}

While the little Higgs models are likely to result in a very limited phenomenology at the Tevatron energies, the first attempt to look for additional gauge bosons predicted in these models has been made by the CDF Collaboration. The analysis is a straightforward extension of the search for Randall-Sundrum gravitons in the dielectron and dimuon channels, with the cross section limits on a hypothetical narrow resonance signal interpreted as limits on $Z_H$, an additional heavy $SU(2)$ gauge boson predicted in little Higgs models. The parameters that fix the couplings of the $Z_H$ to the SM fields are discussed in Ref.~\cite{LH}; the two relevant ones are the mass of the $Z_H$ and the parameter $\cot\theta$ that governs $Z_H$ mixing with other neutral gauge bosons. Figure~\ref{fig:LH} shows the 95\% CL exclusion contours obtained in each channel. This is the first search for $Z_H$ at colliders.

\section{CONCLUSIONS}

Searches for low-energy effects predicted in recent models with extra spatial dimensions are among the most exciting aspects of geenral program of searches for physics beyond the SM. Various existing constraints on several ED models are reviewed and the latest results from the Tevatron Run II are presented. In most of the cases they yield the most stringent limits on the model parameters to date and thus have reached a new level of sensitivity to the effects of ED. With the Tevatron performance exceeding its design level and both the CDF and D\O\ experiments running efficiently, these searches will continue at high pace and move into an uncharted territory in the allowed model parameter space. While the LHC is the ultimate machine where new models with extra dimensions will be either confirmed or closed, in the next few years the Tevatron has an exciting opportunity to either catch first glimpse of ED or to expand significantly the excluded parameter space. An exciting discovery might be just around the corner!

\begin{acknowledgments}
I would like to thanks the organizers of the SSI `04 for a kind invitation and for an excellent and productive Institute. I am grateful to my colleagues in the CDF and D\O\ Collaborations for providing experimental results for this talk. Special thanks to Volker Buescher, Ray Culbertson, Laurent Duflot, Jean-Francois Grivaz, and Beate Heinemann. This work is supported by the Alfred P. Sloan Foundation, the National Science Foundation under the CAREER Award PHY-0239367, and the U.S.~Department of Energy under Grant No. DE-FG02-91ER40688. Last but not the least, I am indebted to Natalie for her devotion and particularly for putting up with my busy summer travel schedule.
\end{acknowledgments}


\end{document}